\documentclass[12pt]{JHEP3}

\usepackage{epsfig}
\epsfclipon

\def\hf{{1\over 2}}
\def\Dsl{\hbox{/\kern-.6700em\it D}} 
\def\dsl{\hbox{/\kern-.5300em$\partial$}}

\def\eps{\epsilon}

\def\Q{{\cal Q}}

\def\exd{\hbox{d}}
\def\sss{\scriptscriptstyle}

\newcommand{\sfrac}[2]{{\textstyle\frac{#1}{#2}}}
\def\eq{\begin{equation}}
\def\eeq{\end{equation}}
\def\eqa{\begin{eqnarray}}
\def\eeqa{\end{eqnarray}}
\def\nn{\nonumber}

\def\bd{\begin{displaymath}}
\def\ed{\end{diplaymath}}

\def\Box{ {\,\lower 0.9pt\vbox{\hrule\hbox{\vrule height0.2cm \hskip 0.2cm \vrule height 0.2cm }\hrule}\,}}
\def\lsim{{\ \lower-1.2pt\vbox{\hbox{\rlap{$<$}\lower5pt\vbox{\hbox{$\sim$}}}}\ }}
\def\gsim{{\ \lower-1.2pt\vbox{\hbox{\rlap{$>$}\lower5pt\vbox{\hbox{$\sim$}}}}\ }}

\def\pref#1{(\ref{#1})}

\def\ssubsubsection#1{\vspace{3mm} \noindent \textbf{#1} \\ \vspace{-3mm} \\ \noindent}

\def\cA{{\cal A}}
\def\cF{{\cal F}}

\def\cL{{\cal L}}

\def\hf{{1\over 2}}

\def\Dsl{\hbox{/\kern-.6700em\it D}} 
\def\dsl{\hbox{/\kern-.5300em$\partial$}}
\def\veps{\varepsilon}
\def\eps{\epsilon}

\def\nn{\nonumber}

\def\implies{\Longrightarrow}

\def\implies{{\Rightarrow}}

\def\beginvector{\left( \begin{array}{c} }
\def\endvector{\end{array} \right)}
\def\endignore{}
\def\ignore#1\endignore{}

\def\eps{\epsilon}
\def\di{\partial}

\def\dii#1#2{{\partial #1 \over \partial #2}}
\def\S2{{\mathcal S}^2}

\def\myref#1{(\ref{#1})}

\title{Warped Brane Worlds in \\ Six Dimensional Supergravity}
\author{Y. Aghababaie,$^{1}$ C.P. Burgess,$^{1}$, J.M. Cline,$^{1,2}$
H. Firouzjahi,$^{1}$ S.L. Parameswaran,$^3$ F. Quevedo,$^{3}$ G.
Tasinato $^{4}$ and I. Zavala $^{3}$
\\

$^1$ Physics Department, McGill University,
                3600 University Street,\\
                Montr\'eal, Qu\'ebec, Canada, H3A 2T8.\\

$^2$ Theory Division, CERN, Gen\`eve 23, Switzerland.\\

$^3$ Centre for Mathematical Sciences, DAMTP,
               University of Cambridge,\\
               Cambridge CB3 0WA UK.\\

$^4$ Physikalisches Institut der Universit\"at Bonn, \\
    Nussallee 12, 53115 Bonn, Germany.}

\abstract{We present warped compactification solutions of six-dimensional
supergravity, which are generalizations of the
Randall-Sundrum (RS) warped brane world to codimension two and to a
supersymmetric context. In these solutions the dilaton varies over
the extra dimensions, and this makes the electroweak hierarchy
only power-law sensitive to the proper radius of the extra
dimensions (as opposed to being exponentially sensitive as in the
RS model). Warping changes the phenomenology of these models
because the Kaluza-Klein gap can be much larger than the internal
space's inverse proper radius. We provide examples both for
Romans' nonchiral supergravity and Salam-Sezgin chiral
supergravity, and in both cases the solutions break all of the
supersymmetries of the models. We interpret the solution as
describing the fields sourced by a 3-brane and a boundary 4-brane
(Romans' supergravity) or by one or two 3-branes (Salam-Sezgin
supergravity), and we identify the topological constraints which
are required by this interpretation. For both types of solutions
the 3-branes are flat for all topologically-allowed values of the
brane tensions. We identify the general mechanism for and
limitations of the self-tuning of the effective 4D cosmological
constant in higher-dimensional supergravity which these models
illustrate.}

\preprint{McGill-03/14, DAMTP-2003-77, CERN-TH/2003-190}

\keywords{brane-world, supergravity, warped geometries}

\begin{document}

\section{Introduction}
Although higher-dimensional models have a long history within
supersymmetric theories, there has been less exploration within
supergravity theories of the low-energy implications of warped
compactifications \cite{5Dsugrawarp, berglund}. This is by contrast with
nonsupersymmetric models, for which warped compactifications
have been explored in some detail in 5 spacetime dimensions
\cite{RS}, and more recently in 6 dimensions
\cite{AdSsoliton,6DRS,OurADS}.  The reason for this difference is
partly due to the point of view taken by workers on the 5D models,
for whom part of the basic motivation was to provide an approach
to the hierarchy problem which is an alternative to supersymmetric
models.

In the end Nature may not feel the need to choose to solve the
hierarchy problem using only supersymmetry or only warping.
Warping may play a role in the hierarchy problem in addition to
supersymmetry, rather than in competition with it. In order to
decide whether or not it does requires a better theoretical
exploration of what is possible. Certainly if string theory proves
to be the correct theory of very short distances warping is only
likely to play a role at low-energies within a supergravity
framework.

One of the main difficulties to constructing warped brane-world
models in general is the absence of explicit solutions describing
branes within compact spaces including the back-reaction on the
space due to the branes. The Randall-Sundrum \cite{RS}
construction provides such solutions for 3-branes in five
dimensions, with the transverse dimension described by a line
segment. (Solutions to the corresponding 5D
supergravity equations are also known
having a warped  geometry \cite{5Dsugrawarp}.) What makes these
solutions possible is the fact that the branes have codimension
1, and so their gravitational back-reaction may be summarized by
the Israel junction conditions.

Six-dimensions are also attractive for constructing brane world
models with compact internal spaces since the gravitational
back-reaction problem for 3-branes (codimension two objects) is
also soluble in terms of $\delta$-function curvature singularities
\cite{Deser}. Warped examples of this type have been constructed
\cite{6DRS,OurADS}, based on the AdS soliton solution
\cite{AdSsoliton} to the Einstein equations with negative
cosmological constant. Unwarped brane-world solutions have also
been constructed, both for nonsupersymmetric \cite{football} and
for supersymmetric \cite{branesphere} systems \footnote{Codimension two
warped solutions of type IIB string theory have also been considered \cite{berglund},
with supersymmetry broken by a global cosmic brane of finite extent.}.  

In this paper we describe the first examples of warped brane-world
compactifications of six-dimensional supergravity. We do so by
explicitly solving the 6D coupled Einstein-Maxwell-dilaton
equations in both their Romans' \cite{RomansSG,NPST} and
Salam-Sezgin \cite{MS,NS,SS} variants
\footnote{These solutions are analytical continuations of the solutions
recently found in  \cite{FatPaper}. While writing this article a general
solution of the Salam-Sezgin supergravity
was discussed in \cite{ggp} having similar properties as the ones discussed
in section 3.}.  In all of these solutions
the warping of the 4D metric goes hand in hand with a nontrivial
dilaton configuration, and so these solutions generalize the
simpler product-space spherical compactifications of the
Salam-Sezgin model \cite{SS,RSSS,susysphere,branesphere,garychris}. Unlike the
spacetime curvature, the dilaton and electromagnetic fields in our
solution {\it ansatz} are nonsingular at the positions of the
3-branes, and so the solutions can only describe the fields due to
3-branes which do not couple to these fields.

Typically we find that the warped solutions for Romans'
supergravity resemble and generalize the nonsupersymmetric AdS
soliton solutions, with the internal dimensions being bounded by a
4-brane and containing `our' 3-brane at an interior point. By
contrast, the solutions for Salam-Sezgin supergravity generalize
the unwarped solutions, for which a pair of 3-branes sit at
opposite poles of an internal 2-sphere. In neither case does the
electroweak hierarchy depend exponentially on the size of the
internal dimensions, because the solutions are not asymptotically
anti-de Sitter. For Salam-Sezgin supergravity they are not because
the scalar potential is positive. For Romans' supergravity it is
because the dilaton field varies in such a way as to run
asymptotically to a zero of the potential. Because it is not
exponential we find that some dimensionless combinations of brane
tensions and couplings must be chosen to be very large if the
hierarchy is to be sufficiently big.

In all of our solutions the internal geometry of the 3-branes is
flat for all values of the brane tensions. This provides a final
motivation for exploring their properties: to explore further the
nature of the self-tuning of the effective 4D cosmological
constant in 6D supergravity theories, as was discussed for
unwarped compactifications in ref.~\cite{branesphere}. We find the
result that self-tuning also occurs for these warped
compactifications provided the branes are assumed to have specific
kinds of charges, but without requiring any adjustments of the
bulk coupling constants.

We organize our presentation as follows. The next section
describes the warped solution to Romans' supergravity, and in
particular examines how its low-energy features (such as the
electroweak hierarchy) depend on the physical properties of the
branes involved. Section 3 repeats this discussion for warped
compactifications of Salam-Sezgin supergravity. The nature of the
cosmological-constant self-tuning is then described for both
models in section 4. Finally, our conclusions are summarized in
section 5.

\section{Romans Supergravity in 6D}
We now describe nonchiral six-dimensional supergravity, which is
the first example for which we present warped compactifications.
The solutions which we find in this case are supersymmetric
generalizations of the well-known AdS soliton \cite{AdSsoliton}.
The 6D supergravities described in this section are the $N=4^g$
and $N=\tilde{4}^g$ models of ref.~\cite{RomansSG}. It is known
how to obtain these theories from 10D supergravity, and we
summarize in an appendix how the solutions we find may be lifted,
or oxidized, to higher dimensions than six.

\subsection{The Model}
The bosonic field content of the theory consists of a metric
($g_{MN}$), antisymmetric gauge field ($B_{MN}$), dilaton ($\phi$)
plus 6D gauge potentials ($A^\alpha_M$) for the gauge group $G = SU(2)
\times U(1)$. The fermionic field content comprises 4 gravitini
($\psi^i_M$) and four spin-$\frac12$ fields ($\chi^i$), where $i =
1,2$ are indices on which the $SU(2)$ gauge-group factor acts with
generators ${(T_\alpha)^i}_j$. (We use the index $I = 1,\dots,4$
to denote $G$ generators and $\alpha = 1,2,3$ to label the $SU(2)$
subgroup.) The fermions all satisfy an $SU(2)$ symplectic Majorana
condition. Although this condition is compatible with a
simultaneous Weyl condition in six dimensions, we do {\it not}
impose this additional condition, and so the theory is trivially
anomaly free. For instance, the fermion covariant derivative is
\begin{equation} \label{E:covderivR}
    D_{M}\chi^i = \left[ \left( \partial_{M} +
    \frac{1}{4}{\omega_{M}}^{AB}\Gamma_{AB} \right) {\delta^i}_j
    + g_2 A^\alpha_{M} \, {(T_\alpha)^i}_j \right]\chi^j_{N} ,
\end{equation}
where ${\omega_{M}}^{AB}$ denotes the spin connection and $g_2$
denotes the 6D $SU(2)$ gauge coupling. The field strengths for
$B_{MN}$ and the $U(1)$ gauge potential, $\cA_M$, are the usual
abelian expressions $G = \exd B$ and $\cF = \exd \cA$, while
$F^\alpha_{MN}$ denotes the usual $SU(2)$ nonabelian field
strength.

The bosonic part of the classical 6D supergravity action
is:\footnote{Our conventions differ from ref.~\cite{RomansSG} in
that we use Weinberg's curvature conventions \cite{GandC} and we
set $\kappa_6^2 = 8 \pi G_6 = 1$ rather than 2, and with this
choice the canonical normalization of the dilaton requires $\phi_R
\to \phi = \sqrt{2} \, \phi_R$.}
\eqa \label{E:BactionR}
    e_6^{-1} {\cal L}_B &=& -\, \frac{1}{2 } \, R - \frac{1}{2 } \,
    \partial_{M} \phi \, \partial^M\phi  -
    \frac{1}{12} \, e^{-2\zeta \phi} \; G_{MNP}G^{MNP} \nn\\
    && \qquad \qquad  + \frac12 \, {g^2_2 \,  e^\phi}
    -\, \frac{1}{4} \, e^{-\phi} \; \Bigl(F^\alpha_{MN}F_\alpha^{MN}
    + \cF_{MN} \cF^{MN} \Bigr) \nn\\
    && \qquad \qquad   - \,
    \frac{1}{8 \sqrt{2}} \, \eps^{MNPQRS} B_{MN} \Bigl( F^\alpha_{PQ} F_{\alpha RS}
    + \cF_{PQ} \cF_{RS} \Bigr),
\eeqa
where as usual $e_6 = |\det {e_M}^A| = \sqrt{-\det g_{MN}}$. The
parameter $\zeta$ which defines the dilaton coupling to $G_{MNP}$
takes values $\zeta = -1$ for Romans' $N = 4^g$ theory. By
contrast it is $\zeta = +1$ for the $N = \tilde{4}^g$ theory,
which is obtained from $N = 4^g$ by dualising $G_{MNP} \to
\widetilde{G}_{MNP} = \frac{1}{3!} \, e^{-2 \phi} \, \eps_{MNPQRS}
\, G^{QRS}$.

For later purposes it is useful to record here the supersymmetry
transformation rules for the fermions of the model. For the $N =
4^g$ theory these are
\eqa \label{E:Romansusy}
    \delta \chi^i &=& \frac{1}{\sqrt2} \, \Gamma^M  \epsilon^i
    \, \partial_M \phi +
    \frac{g_2\, e^{\phi/2} }{2\sqrt{2}} \, \Gamma_7 \epsilon^i -
    \frac{e^{\phi}}{12}  \Gamma_7 \Gamma^{MNP} \epsilon^i
    G_{MNP} \nn \\
    && \qquad + \frac{e^{-\phi/2}}{4\sqrt{2}} \, \Gamma^{MN}
    \Bigl(\cF_{MN} \delta_j^i + 2\, \Gamma_7
    F^\alpha_{MN} (T_\alpha)_j^i \Bigr) \, \epsilon^j \\
    \delta \psi_M^i &=& \sqrt2 \, D_M \epsilon^i - \frac{g_2
    \, e^{\phi/2}}{4
    \sqrt{2}}  \, \Gamma_M \Gamma_7 \epsilon^i
    - \frac{ e^{\phi} }{24} \,\Gamma_7 \Gamma^{PQR}
    \epsilon^i \, G_{PQR} \nn\\
    &&\qquad  - \frac{e^{-\phi/2}}{8 \sqrt{2}} \Bigl(
    {\Gamma_M}^{PQ} - 6 \delta_M^P \Gamma^Q \Bigr)  \,
    \Bigl( \cF_{PQ} \delta^i_j + 2 \, \Gamma_7 F^\alpha_{PQ}
    (T_\alpha)^i_j \Bigr) \, \epsilon^j \, ,
\eeqa
where $\epsilon$ is the supersymmetry parameter. As usual
\eq
    \Gamma^{A_1\dots A_n} = \frac{1}{n!} \, \left[
    \Gamma^{A_1}\cdots \Gamma^{A_n} \pm \hbox{permutations} \right]
\eeq
denotes the completely antisymmetric product. Letters from the
beginning of the alphabet denote tangent-frame indices, those from
the middle of the alphabet denote world indices, and these are
related to one another by the vielbein by $\Gamma^M = {e_A}^M \,
\Gamma^A$. The connection appearing in the covariant derivative of
the SUSY parameter $\epsilon^i$ here is given by the spin
connection, according to
\begin{equation}
    \nabla_{M} \epsilon^i \equiv\left( \partial_{M}+\frac{1}{4}
    \omega_{M}^{AB} \Gamma_{AB} \right)
    \epsilon^i \,,
\end{equation}
and in our conventions $\Gamma_7^2 = 1$.

\subsection{Warped Compactifications}
We now turn to the 4D compactifications of the model, for which
the internal two dimensions are rotationally invariant about the
position of a centrally-placed 3-brane. In order to make the
resulting two dimensions compact we also take the outer edge of
this two-dimensional space to be bounded by a 4-brane.

The field equations which follow from the action,
eq.~\pref{E:BactionR}, are
\eqa \label{E:Beom}
    &&\Box \, \phi + \frac{\zeta}{6} \, e^{-2 \zeta  \phi}
    \, G^2 + \frac14 \, e^{-\phi} \,
    \Bigl(  F^2 + \cF^2
    \Bigr) + \frac{g_2^2}{2} \, e^\phi = 0 \nn\\
    &&D_P \Bigl( e^{-2 \zeta \phi} \, G^{PMN} \Bigr)  - \,
    \frac{1}{4 \sqrt{2}} \, \eps^{MNPQRS} \Bigl( F^\alpha_{PQ} F_{\alpha RS}
    + \cF_{PQ} \cF_{RS} \Bigr)= 0  \\
    &&D_M \Bigl( e^{-\phi} \, F^{MN}_\alpha \Bigr) -\, \frac{1}{6\sqrt2} \,
    \eps^{RSPQMN} G_{MRS} \, F_{\alpha PQ} = 0 \nn \\
    &&D_M \Bigl( e^{-\phi} \, \cF^{MN} \Bigr) -\, \frac{1}{6\sqrt2} \,
    \eps^{RSPQMN} G_{MRS} \, \cF_{PQ} = 0 \nn \\
    &&R_{MN} + \partial_M\phi \, \partial_N \phi +
    \frac12 \, e^{-2 \zeta \phi} \, G_{MPQ} \, {G_N}^{PQ}
    + \, e^{-\phi} \, \Bigl( F_{MP}^\alpha
    {F_{\alpha N}}^P + \cF_{MP} {\cF_N}^P \Bigr) \nn\\
    &&
    -\, \left[  \frac{1}{12} \, e^{-2 \zeta  \phi}
    \, G^2 + \frac18 \,  e^{-\phi} \, \Bigl( F^2 + \cF^2 \Bigr)
    + \frac{g_2^2}{4} \, e^\phi \right]  \,  g_{MN} = 0 , \nn
\eeqa
Notice that when $B_{MN} = 0$ these equations are invariant under
the rescaling $g_{MN} \to \Omega g_{MN}$, $e^\phi \to \Omega^{-1}
e^\phi$, for constant $\Omega$.

To obtain solutions we adopt the following {\it ans\"atze}:
\eqa \label{ansatze}
    ds_6^2 &=& a(r) \Bigl[ h_{\mu\nu}(x) \, \exd x^\mu \, \exd x^\nu +
    b(r) \, \exd \theta^2 \Bigr] + \frac{\exd r^2}{a(r) \, b(r)}
    \nn\\
    F^{\hat\alpha}_{r\theta} &=& f(r) \, \eps_{r\theta} \\
    \phi &=& \phi(r) \, , \nn
\eeqa
where $\epsilon_{r\theta} = \pm e_2$ is the volume form in the
internal two dimensions, and $\hat\alpha$ denotes the gauge-group
element whose background field strength is nonzero. All other
fields vanish. The intrinsic 4D metric, $h_{\mu\nu}$ is assumed to
be maximally symmetric but warped, with warp factor $a(r)$. This
is the most general form which is consistent with the product of
maximal symmetry in the four noncompact dimensions and rotational
invariance in the two internal dimensions. In practice our
interest is particularly in solutions for which the intrinsic four
dimensions are flat: $h_{\mu\nu} = \eta_{\mu\nu}$, and in whether
this requires a fine-tuning of the theory's couplings.

With these choices the generalized Maxwell equation becomes
\eq \label{Maxwell}
    \Bigl( e_6 \, e^{-\phi} \, F^{r\theta}_{\hat\alpha} \Bigr)' =
    \Bigl(a^2 \,{e^{-\phi} \, f} \Bigr)' = 0 \, ,
\eeq
where the prime denotes differentiation with respect to the
coordinate $r$. This has as solution
\eq \label{fresult}
    f(r) = {A \, g_2 \, e^\phi \over a^2} \, ,
\eeq
where $A$ is an arbitrary constant of integration and the factor
$g_2$ is included for later convenience.

This expression for $f$ is sufficient to exclude the possibility
of obtaining solutions with constant $\phi$, as may be seen by
using the above {\it ansatz} with $\phi' = 0$ in the dilaton field
equation, leading to
\eq
    0 = f^2 + g_2^2 \, e^{2\phi} = g_2^2 \, e^{2\phi} \left(1 +
    \frac{A^2}{a^2} \right) \, .
\eeq
Clearly this cannot be satisfied for real fields and nonzero $g_2$
and $A$ unless $\phi \to -\infty$.

Using eq.~\pref{fresult}, the Einstein equations reduce to the
system
\eqa \label{Einstein}
    \frac{a''}{a} + 2 \, \left( \frac{a'}{a} \right)^2 + \frac{a'
    \, b'}{a b} &=& \frac{g_2^2 \, e^\phi}{2 a b} \left(1 + \frac{
    A^2}{a^4} \right) \nn\\
    \frac{a''}{a} + \frac{b''}{b} + \frac{4 \, a' b'}{a b} + 2 \left(
    \frac{a'}{a} \right)^2 &=& \frac{g_2^2 \, e^\phi}{2 a b}
    \left(1  - \frac{3 A^2 }{a^4} \right) \\
    \frac{5 \, a''}{a} + \frac{b''}{b} +  \frac{4 \, a' b'}{a b} &=& -2
    \, \phi'^{\,2} + \frac{g_2^2\, e^\phi}{2 a b} \left(1  - \frac{3
    A^2 }{a^4} \right) \, . \nn
\eeqa
We need not explicitly write the dilaton equation, as this is not
independent of the ones already written so long as $\phi' \ne 0$.

To find solutions it is useful to eliminate $b$ by taking the
difference of the last two of eqs.~\pref{Einstein}, to obtain
\eq \label{MEq1}
    2 \left( \frac{a''}{a} \right) - \left( \frac{a'}{a} \right)^2
    = - \, \phi'^{\, 2} \, .
\eeq
It is also useful to eliminate all second derivatives by
subtracting half of this last equation from the first of
eqs.~\pref{Einstein}, giving
\eq \label{MEq2}
    5 \left( \frac{a'}{a} \right)^2 + 2 \left( \frac{a' b'}{a b}
    \right) = \phi'^{\, 2} + \frac{g_2^2 \, e^\phi}{ a b} \left(
    1  + \frac{A^2}{a^4} \right) \, .
\eeq
The invariance of eq.~\pref{MEq1} with respect to rescaling $r$
suggests a power-law solution,
\eq \label{phiasoln}
    \phi(r) = \phi_0 - \sqrt{p \, (2 - p)} \,  \ln r \qquad
    a(r) = a_0 \, r^{p} \, ,
\eeq
where $0 < p < 2$. Notice the cases $p = 0$ and $p = 2$ may be
excluded because these would imply $\phi' = 0$, which we have seen
is inconsistent with the dilaton field equation.

Using this in eq.~\pref{MEq2} gives a linear first-order equation
for $b(r)$, whose general solution is
\eq \label{bsolngen}
    b(r) = b_1 \, r^{\beta_1} - b_2 r^{\beta_2} - B \,r^{\beta_h} \, , \eeq
where $B$ is an integration constant, while
\eq \label{bpowers}
    \beta_1 = 2 - p - \sqrt{p(2-p)} \, , \qquad
    \beta_2 = 2 - 5p - \sqrt{p(2-p)} \, , \qquad
    \beta_h = 1 - 3p \, ,
\eeq
and
\eq \label{bcoefs}
    b_1 = \frac{g_2^2 \, e^{\phi_0}}{2p \, a_0 \, (\beta_1
    + 3p - 1)} \, , \qquad
    b_2 = \frac{A^2 \, g_2^2 \, e^{\phi_0}}{2p \,a_0^5 (1 - 3p -
    \beta_2)} \, .
\eeq

Finally, substitution of this solution back into
eqs.~\pref{Einstein} (or into the dilaton equation) shows that we
must further require $p = 1$. This leaves the final expression
\eq \label{bsoln}
    b(r) = b_1  - \frac{B}{r^{2}} - \frac{b_2}{r^{4}} \, , \eeq
with
\eq
    b_1 = \frac{g_2^2 \, e^{\phi_0}}{4 \, a_0} \, , \qquad
    b_2 = \frac{A^2 \, g_2^2 \, e^{\phi_0}}{4\, a_0^5} \, ,
\eeq
both positive.

In summary, we obtain in this way as solutions the explicit field
configurations
\eq
    \phi(r) = \phi_0 - \,  \ln r \qquad
    f(r) = {A  \, g_2\, e^{\phi_0} \over a_0^2 \, r^3} \qquad
    a(r) = a_0 \, r  \qquad
    b(r) = b_1   -\, \frac{B}{r^2} - \,  \frac{b_2}{r^4}\, .
\eeq
This solution is also obtainable by appropriately continuing the
solutions of ref.~\cite{FatPaper}.

For future reference we also note that the nonvanishing component
of the gauge potential itself is given locally by
\eq
\label{gaugepot}
    A_\theta^{\hat\alpha} = C - {A\, g_2\, e^{\phi_0}\over 2 a_0^2  r^2}
\eeq
where, as before, $\hat\alpha$ is the $SU(2)\times U(1)$ index of the
background field and $C$ is a constant of integration.

At first sight there appear to be 5 constants of integration to be
determined: $A,B,C, a_0, \phi_0$. Physically, $A$ corresponds to
the strength of the gauge-field flux, and $B$ is analogous
to the black hole mass in the Schwarzschild solution. Locally, $C$
is an irrelevant gauge degree of freedom, but we keep it here
since it can encode gauge-invariant information in spaces with
nontrivial topology.

In fact, only 4 of these 5 are set by boundary conditions because
the rescaling symmetry of the field equations for $B_{MN}=0$ allow
one combination to be set by an appropriate choice of units in the
4 noncompact dimensions. To see this explicitly, perform the
following rescaling of the integration constants $A,B, \phi_0,
a_0$:
\eqa \label{rescaling}
e^{\phi_0}\to  e^{\phi_0} &  \qquad &  a_0\to c^{-2} a_0 \nn \\
A \to c^{-4} A & \qquad & B\rightarrow c^2 B \,,
\eeqa
where $c$ is an arbitrary constant parameter. This rescaling does
not alter the form of the solution obtained above, because its
effects can be compensated by performing the coordinate
transformation $x^\mu\to c x^\mu$, with $r$ and $\theta$ held
fixed. Below we will use this rescaling to fix $a_0$, and so
to reduce the number of integration constants to 4. For the
moment, however, we keep all 5 parameters.

Notice also that even once $a_0$ is fixed in this way, the
rescalings
\eq \label{rescaling2}
    r \to \Omega \, r, \qquad B \to \Omega^2 \, B, \qquad
    A \to \Omega^2 \, A \,,
\eeq
have the effect of rescaling the solution according to $g_{MN} \to
\Omega \, g_{MN}$, $e^\phi \to \Omega ^{-1} \, e^\phi$ and $F_{MN}
\to F_{MN}$, which we have seen is a symmetry of the classical
equations. To the extent that the boundary conditions also respect
this symmetry we should not expect to be able to determine the
combination of integration constants which corresponds to this
rescaling.

Finally, notice that the solution only depends on the gauge
coupling, $g_2$, and $\phi_0$ through the combination $g_2^2 \,
e^{\phi_0}$. As such, one can --- although we shall not --- set
$g_2 = 1$ during all manipulations, secure in the knowledge that
the appropriate factors of $g_2$ can be restored easily.

\subsubsection{Supersymmetry of the Solution}
We examine the supersymmetry of this solution, and show that
it is supersymmetric only if $A=B=0$. To do so we evaluate the
supersymmetry transformations, \pref{E:Romansusy}, at the bosonic
solution with vanishing fermion fields, $G_{\it 3}= 0$, and with a
single $U(1)$ gauge field. With these choices we are left with the
equation\footnote{We are 
 outlining the calculation for the case where the vacuum expectation value of the gauge field
lies in the $U(1)$ subgroup, but  the result is the same in the case that it lies
in  $SU(2)$.}
\eq\label{gaugino}
    \delta \chi^i = \frac{1}{\sqrt2} \, \Gamma^M  \epsilon^i
    \, \partial_M \phi +
    \frac{g_2\, e^{\phi/2} }{2\sqrt{2}} \, \Gamma_7 \epsilon^i
     + \frac{e^{-\phi/2}}{4\sqrt{2}} \, \Gamma^{MN}
    \, \epsilon^i \, F_{MN}  \, ,
\eeq
while for the gravitino we have
\eq\label{gravitino}
    \delta \psi_M^i = \sqrt2 \, D_M \epsilon^i - \frac{g_2
    \, e^{\phi/2}}{4
    \sqrt{2}}  \, \Gamma_M \Gamma_7 \epsilon^i
     - \frac{e^{-\phi/2}}{8 \sqrt{2}} \Bigl(
    {\Gamma_M}^{PQ} - 6 \delta_M^P \Gamma^Q \Bigr)  \,
    \, \epsilon^i\, F_{PQ} \, .
\eeq

Let us concentrate on the condition $\delta \chi_{i}=0$ since the
condition found using the gravitino transformation can be worked
out in a similar manner. Specializing as above we
have
\eq\label{susystep1}
    \Gamma^{r}\phi'
    \,\epsilon+\frac{g_2}{2} e^{\phi/2} \Gamma_{7}\, \epsilon
    +\frac{1}{2} e^{-\phi/2} \Gamma^{r}\Gamma^{\theta}
    \, F_{r \theta}\, \epsilon=0\,.
\eeq
Multiplying both sides of this equation by the tangent-frame Dirac
matrix in the $r$ direction, $\overline{\Gamma}^r =
\Gamma^r/\sqrt{g^{rr}}$, and considering the two possible
eigenvalues
\begin{equation}
    \overline{\Gamma}^{r} \Gamma_{7} \, \epsilon_{\pm} = \pm \epsilon_{\pm}\,,
\end{equation}
the equation (\ref{susystep1}) becomes
\eq\label{susystep2}
    \left[ \sqrt{g^{rr}} \phi' \pm \frac{g_2}{2} e^{\phi/2} \right] \epsilon_{\pm}
    =-\frac{1}{2} e^{-\phi/2} \sqrt{g^{rr}} \Gamma^{\theta} \, F_{r\theta}
    \, \epsilon_{\pm} \,.
\eeq
Squaring both sides of this equality then allows us to remove all
Dirac matrices, leading to the equation
\eq\label{susystep3} \left[ \sqrt{g^{rr}} \phi' \pm \frac{g_2}{2}
e^{\phi/2} \right]^{2}\, \epsilon_{\pm} = \frac{1}{4} e^{-\phi}
g^{rr}g^{\theta\theta} F_{r\theta}^2\, \epsilon_{\pm}
\eeq
which is satisfied for our solution only if both sides vanish,
requiring $A=B=0$.

\subsubsection{Conical Singularities}
The metric which results from these functions describes a geometry
which is singular for $r \to 0$, where there are curvature
invariants which diverge. Because we regard our field equations
as valid only in the limit of small curvatures, we cannot trust
our solution in this region. Furthermore, since both $b_1$ and
$b_2$ are nonnegative the function $b(r)$ is negative for $r \to
0$ and positive for $r \to \infty$, passing through zero at the
point $r = r_3$, where\footnote{If $b_2=0$, such as when $A=0$,
then instead one finds $r_3^2 = B/b_1$.}
\eq \label{rthree}
    \frac{2 b_2}{r_3^2} = -B + \sqrt{ B^2 + 4 b_1 b_2} \, .
\eeq
The metric therefore has Lorentzian signature for $r > r_3$, while
for $r < r_3$ its signature is $(3,3)$.

Since the proper circumference of a circle at fixed $r$ goes to
zero as $r \to r_3$, the Lorentzian-signature space ($r \ge r_3$)
pinches off there, and although all curvature invariants have
smooth limits as $r \to r_3$ the metric acquires a conical
singularity at this point. The conical singularity is exhibited by
writing $r = r_3 + \delta$ (with $\delta \ll 1$), in which case
the two-dimensional metric becomes
\eqa
    ds_2^2 &=& a(r)\, b(r) \, \exd \theta^2 +
    \frac{\exd r^2}{a(r) \, b(r)} \nn\\
    &\approx& a_3\, b_3' \delta \, \exd \theta^2 +
    \frac{\exd \delta^2}{a_3 \, b_3' \delta}  \\
    &=& \left(\frac{1}{a_3 \, b_3'} \right) \left[ \exd \rho^2 +
    \left( \frac{a_3 \, b_3'}{2} \right)^2 \rho^2 \exd \theta^2 \right]
    \, , \nn
\eeqa
where $\rho = 2  \sqrt\delta$, $a_3 = a(r_3) = a_0 \, r_3$ and
\eq
    b_3' = \left( \frac{db}{dr}\right)_{r = r_3} = \frac{2B}{r_3^3}
    + \frac{4 b_2}{r_3^5} \, .
\eeq
The geometry is therefore locally a cone, with a delta-function
singularity in the curvature which is proportional to the defect
angle $\Delta \theta = 2\pi \veps_3$, with
\eq \label{deficit1}
    \veps_3 = 1 - \, \frac{a_3 b_3'}{2}
    = 1 - a_3
    \left( \frac{B}{r_3^3} + \frac{2 b_2}{r_3^5}
    \right) .
\eeq

If $a_3 b_3'=2$ the solution is nonsingular. This can be achieved
by appropriately restricting the parameters $A,B,a_0,\phi_0$ (such
as by choosing $A=0$ and $g_2^2 \, e^{\phi_0}=4$). Otherwise the
solution has a conical singularity at $r = r_3$, and the
nonvanishing defect angle can be interpreted as the response of
the geometry to the presence of a 3-brane located there.

At this point we use the freedom described earlier to rescale the
4D coordinates to set $a_0 = 1/r_3$, and so to ensure $a_3 = a_0
\, r_3 = 1$. (This is accomplished by choosing $c^2=1/r_3$ in the
scaling transformations (\ref{rescaling}).) It is also convenient
to define a new parameter $\alpha_3 = e^{\phi_0}/r_3$ so that
$e^{\phi(r_3)} = \alpha_3$ denotes the effective 4D bulk gauge
coupling at $r = r_3$. The solution and its dependent parameters
then take the form
\eqa \label{finalform}
    &&e^{\phi(r)} = \alpha_3 \, \left( {r_3\over r} \right) ;\quad
    A^{\hat\alpha}_\theta(r) = C - \frac{A \, g_2 \, \alpha_3 \, r_3^3 }{ 2 \,r^2};
    \nn\\
    &&a(r) = {r\over r_3};  \quad
    b_1 = {g_2^2 \, \alpha_3 \,r_3^2\over 4 }; \quad b_2 =
    {A^2 \, g_2^2 \, \alpha_3 \,r_3^6\over 4 } \, ,
\eeqa
with four independent integration constants, $A,B,C$ and
$\alpha_3$. Since these equations imply that $b_1$ and $b_2$ are
themselves functions of $r_3$, eq.~\pref{rthree} can itself be
solved to give the more explicit result
\eq \label{rthree2}
    r_3^4 = \left( \frac{4B}{g_2^2\, \alpha_3} \right) \, \frac{1}{1 -
    A^2} \, ,
\eeq
from which we see that $B > 0$ implies $|A| < 1$, while $B < 0$
requires $|A| > 1$.

Many of our later expressions simplify considerably in the limits
$A^2 \gg 1$, $A^2 \ll 1$ and $A^2 = 1 - \epsilon$ with $|\epsilon|
\ll 1$, so we list some helpful approximate expressions
here. For this purpose the relation $4 b_1 b_2/B^2 = 4 A^2/(1
- A^2)^2$ is very useful. We have:
\medskip

\begin{itemize}
\item{} {\it The case $A^2 \ll 1$:}

\noindent In this limit we require $B > 0$ and have $r_3^4 \approx
4B/(\alpha_3 g_2^2)$, and so $4 b_1 b_2 /B^2 \approx 4A^2 \ll 1$.
Consequently $r_3^2 \approx B/b_1$ and so $b_2/r^2 < b_2/r_3^2
\approx b_1 b_2/B \ll B$ for all $r > r_3$. This allows the
simplification $b(r) \approx b_1 - B/r^2 = \frac14 \, g_2^2 \,
\alpha_3 \, r_3^2 - B/r^2$ for all $r > r_3$.

\medskip

\item{} {\it The  case  $A^2 \gg 1$:}

\noindent In this limit we must have $B< 0$ and so $r_3^4 \approx
-4B/(A^2\, g_2^2 \, \alpha_3)$. This implies $4 b_1 b_2 /B^2
\approx 4/A^2 \ll 1$. Consequently $r_3^2 \approx -B/b_1$ and so
$b_2/r^2 < b_2/r_3^2 \approx b_1 b_2/|B| \ll |B|$ for all $r >
r_3$. This again allows the simplification $b(r) \approx b_1 -
B/r^2 = \frac14 \, g_2^2 \, \alpha_3 \, r_3^2 - B/r^2$ for all $r
> r_3$.

\medskip

\item{}  {\it The  case  $A^2 = 1 - \epsilon$  with
$|\epsilon| \ll 1$:}

\noindent In this limit we have $r_3^4 \approx 4B/(\alpha_3
\epsilon)$, and so $\hbox{sign}\, B = \hbox{sign}\, \epsilon$.
This implies $4 b_1 b_2 /B^2 \approx 4/\epsilon^2 \gg 1$.
Consequently $r_3^4 \approx b_2/b_1$ and so $b_2/r_3^2 \approx
(b_1 b_2)^{1/2} \gg |B|$. This allows the simplification $b(r)
\approx b_1 - b_2/r^4 = \frac14 \, g_2^2 \, \alpha_3 \, r_3^2(1 -
r_3^4/r^4)$ for  $r_3 < r \lsim (b_2/|B|)^{1/2}$, while $b(r)
\approx b_1 - B/r^2 \approx \frac14 \, g_2^2 \, \alpha_3 \, r_3^2
- B/r^2$ for $r \gsim (b_2/|B|)^{1/2}$.
\end{itemize}
\medskip

We end this section with some relevant expressions characterizing the
geometry of the warped cone.
For this metric the circle with coordinate radius $r_c$ has
circumference
\eq
    \rho(r_c) = 2\pi \, \sqrt{\, a(r_c) \, b(r_c)} = 2\pi \,
    r_3^{-1/2} \Bigl[ b_1 r_c - (B/r_c) - (b_2/r_c^3) \Bigr]^{1/2}
    \, ,
\eeq
and the proper radius of such a circle, measured from the conical
defect, is similarly
\eq
    \ell(r_c) = \int_{r_3}^{r_c} \frac{dr}{\sqrt{ \, a(r) \, b(r)}} =
    {r_3}^{1/2} \,
    \int_{r_3}^{r_c} \frac{dr}{\sqrt{ b_1 r - (B/r) - (b_2/r^3)}} \,
    .
\eeq
If $r_c \gg |B|^{1/2}, b_2^{1/4}$ then $\rho(r_c) \approx 2 \pi \,
\sqrt{b_1 \, r_c/r_3} = \pi g_2 \sqrt{\alpha_3 r_3 r_c}$ and
$\ell(r_c) \approx 2 \sqrt{r_c r_3/ b_1} = (4/g_2) \sqrt{r_c/(r_3
\alpha_3)}$. Notice that these imply the ratio $\rho(r)/\ell(r)$
is independent of $r$ for large enough $r$.

Using the condition $a_3 = 1$, the deficit angle at $r= r_3$,
eq.~\pref{deficit1}, becomes
\eq \label{deficit2}
    \veps_3 = 1 - \, \frac{b_3'}{2}
    = 1 -
    \frac{B}{r_3^3}\left(\frac{1+A^2}{1-A^2}\right) .
\eeq
Clearly this expression always satisfies $\veps_3 \leq 1$, and
$\veps_3 > 0$ implies a lower limit to the ratio $|(A^2-1)/B|$. In
both of the limits $A^2 \ll 1$ and $A^2\gg 1$ the expression for
$\veps_3$ simplifies to $\veps_3 \approx 1-|B|/r_3^3$.

For large $r$ the 2D metric becomes
\eq
    ds_2^2 = a(r) \, b(r) \, d\theta^2 + \frac{dr^2 }{a(r) \, b(r)}
    \approx \frac{b_1 \, r}{r_3} \, d\theta^2 + \frac{r_3 \, dr^2}{b_1 \, r} \, ,
\eeq
which the coordinate transformation $\rho = 2\sqrt{r}$ shows to be
locally flat, but with a conical deficit angle given by
\eq
    \veps_\infty = 1 - \, \frac{b_1}{2 \, r_3}
    = 1 - \, \frac{g_2^2 \, \alpha_3 \, r_3}{8} \, .
\eeq
In general the geometry is one of a cone which is curved near its
apex, and so whose deficit angle differs when measured at infinity
and near the apex: $\veps_\infty \ne \veps_3$.

\subsection{Brane Worlds}
We obtain the desired brane world by placing ourselves on a
3-brane which is located at the position of the conical defect, $r
= r_3$. In this way the conical defect can be ascribed to the
response of the gravitational field to the brane's tension. In
order to obtain a finite extra-dimensional volume the space will
also be terminated at a 4-brane located at $r = r_4$. In this
section we determine how the bulk fields respond to the presence
of these branes, and in so doing relate the integration constants
of the solution just described to the physical properties of the
branes.

\subsubsection{The Electroweak Hierarchy}
The first issue to settle for a brane-world application of these
solutions is how large the space must be in order to properly
describe the electroweak hierarchy $M_w /M_p \sim 10^{-15}$. In
the present case the effective 4D Planck mass may be read off from
the dimensional reduction of the Einstein-Hilbert lagrangian:
\eq
     \int d^2x \, \sqrt{- g_6} \, g^{\mu\nu}
    R_{\mu\nu}(g) = M_p^2 \sqrt{- h} \, \Bigr[ h^{\mu\nu} R_{\mu\nu}(h)
     + \cdots \Bigr] \, ,
\eeq
and so
\eq
    M_p^2 = 2 \pi \, \int_{r_3}^{r_4}
    dr \, a(r) = \frac{\pi}{r_3} (r_4^2 -
    r_3^2) \, .
\eeq
(Recall our units, for which $\kappa_6 = M_6^{-2} = 1$ and $a(r_3)
= 1$.)

This must be compared with the mass scales for particles located
on the 3-brane at $r = r_3$. Taking for example a scalar field,
$\chi$, with mass parameter $\mu_3$, we see that the 3-brane action is
\eq
    \cL_3 = - \, \frac12 \sqrt{-g_4} \Bigl( g^{\mu\nu}
    \partial_\mu \chi \partial_\nu \chi + \mu_3^2 \, \chi^2 \Bigr) =
    - \, \frac12 \sqrt{-h} \, a^2(r_3) \left[
    \frac{h^{\mu\nu}}{a(r_3)} \,
    \partial_\mu \chi \partial_\nu \chi + \mu_3^2 \, \chi^2 \right]
    \, ,
\eeq
and so the particle's physical mass is $m_3 = \mu_3 \sqrt{a(r_3)}
= \mu_3$. For $\mu_3 \sim \kappa_6^{-1/2} = M_6$, $r_3 \ll r_4$,
and dropping $O(1)$ factors we find (after temporarily restoring
powers of $M_6$)
\eq
    \frac{m_3}{M_p} \sim \left( \frac{r_3}{M_6 r_4^2} \right)^{1/2}\,
    .
\eeq

Performing the same exercise for the mass, $m_4$, of particles
confined to the 4-brane at $r = r_4$ gives
\eqa \label{4branemodes}
    \cL_4 &=& - \, \frac12 \sqrt{-g_5} \Bigl( g^{p\, q}
    \partial_p \chi \partial_q \chi + \mu_4^2 \, \chi^2 \Bigr) \nn\\
    &=& - \, \frac12 \sqrt{-h} \, a^{5/2}b^{1/2} \left[
    \frac{h^{\mu\nu}}{a} \, \partial_\mu \chi \partial_\nu \chi +
    \frac{1}{a b} \, (\partial_\theta \chi)^2 + \mu_4^2 \, \chi^2
    \right] \, .
\eeqa
This shows that the Kaluza-Klein (KK) zero mode has a mass of order $m_4 = \mu_4
\, \sqrt{a(r_4)} = \mu_4 \sqrt{r_4/r_3}$, and so if $\mu_4 \sim
\mu_3$, the physical masses satisfy $m_4/m_3 \sim \sqrt{r_4/r_3}$.
Consequently $m_4/M_p \sim ( M_6 r_4)^{-1/2}$.

Eq.~\pref{4branemodes} also implies that the KK masses associated
with the circular direction on the 4-brane have a mass gap $M_4 =
1/\sqrt{b(r_4)}$ --- which becomes $M_4 \approx 1/\sqrt{b_1} =
4/(\hat{g}_2 r_3 \sqrt{\alpha_3})$ for sufficiently large $r_4$.
(Here $\hat{g}_2 = g_2 M_6$ is the dimensionless six-dimensional
gauge coupling.) Notice in particular that the KK mass $M_4$ does
{\it not} vanish in the limit of large $r_4$, although the
relative spacing of KK masses to bare masses as measured purely on
the 4-brane, $M_4/m_4$, does vanish as $r_4 \to \infty$.

In the simplest scenario we choose all parameters except for $r_4$
to be of the same size, and we take the fundamental scale on the 3
brane to be $M_w$: $\mu_3 \sim \mu_4 \sim M_6 \sim 1/g_2 \sim
1/r_3 \sim M_w \sim 1$ TeV. With this choice we have $m_4 \sim
\sqrt{M_w M_p} \sim 3 \times 10^{10}$ GeV, making the scale of the
4-brane the intermediate scale. By contrast, the KK spacing of the
4-brane modes is much smaller (if $g_2^2 \, \alpha_3 \sim 1$),
being of order $1/r_3 \sim M_w$. This implies the intriguing
possibility that massive particles on the 4-brane are naturally
extremely heavy, while the nominally massless modes there form a
KK tower which remains at the electroweak scale.

Since $e^{-\phi}$ pre-multiplies the gauge kinetic terms, $e^\phi$
can be interpreted as a position-dependent modulation of the gauge
coupling. From the form of the dilaton solution we see that the
couplings on the 4-brane are much weaker than those on the
3-brane, by an amount: $e^{\phi(r_4)}/e^{\phi(r_3)} = r_3/r_4 \sim
10^{-15}$. Thus the TeV-mass 4-brane modes are
naturally extremely weakly coupled amongst themselves relative to
the couplings of those TeV modes at $r = r_3$.

With these choices we have $r_4/r_3 \sim (m_4/m_3)^2 \sim
10^{15}$, and so $r_3 \sim (1 \, \hbox{TeV})^{-1} \sim 10^{-19}\,$m
implies $r_4 \sim 0.1\,$mm. This corresponds to proper distance
$\ell(r_4)/r_3 \sim \sqrt{r_4/r_3} \sim 3 \times 10^7$, or
$\ell(r_4) \sim 3 \times 10^{-12} \, \hbox{m} \sim 0.03$
Angstroms. Hence $\ell(r_4)$ is well below the current limits on
short-distance deviations from Newton's Gravitational Law
\cite{NewtonTests}, although this limit is more properly compared
with the scale of KK masses in the bulk.

\subsubsection{Bulk KK Modes}
Decomposing the bulk KK modes as $\Psi(x,r,\theta) = \sum_{nl}
\psi_{nl}(x) \, u_{nl}(r) \, e^{i n \theta}$, the bulk action becomes
\eqa
    \cL_4 &=& \int dr \, d\theta \, \sqrt{-g_6} \Bigl(
    \Psi^* g^{MN} \nabla_M \nabla_{n} \Psi \Bigr) \nn\\
    &=& \sum_{nl}\sqrt{-h} \, \psi^*_{nl} \Bigl( h^{\mu\nu} \, \nabla_\mu
    \nabla_\nu - \lambda_{nl} \Bigr) \psi_{nl} \, ,
\eeqa
which follows from the orthogonality relation
$\int dr \, d\theta \, a \, u^*_{nl} \, u_{mk} e^{i(m-n)\theta} =
\delta_{mn}\, \delta_{lk}$ satisfied by the mode functions,
and the eigenvalue condition
\eq \label{evaleqn}
    \Delta_2 \, u_{nl} = - \frac{1}{a} \Bigl( a^3 \, b \, u_{nl}'
    \Bigr)' + \frac{n^2}{b} \, u_{nl} = \lambda_{nl} \, u_{nl} \, ,
\eeq
where primes denote differentiation with respect to $r$. The
eigenvalues, $\lambda_{nl}$, then give the squares of the bulk KK
masses.

\FIGURE{ \centerline{\epsfxsize=4.in\epsfbox{lnU.eps}}
\label{fig:Uofr}
\caption{\small $\ln |U(r)|$ vs.\ $\ln r/r_3$ for two
illustrative cases. For the solid curve, $U(r) < 0$ near $r =
r_3$, but changes sign as $r$ gets larger. $U(r)$ is always
positive for the dashed curve. } }

We may estimate how the KK masses scale for large $r_4$ by looking
for WKB solutions to eq.~\pref{evaleqn}. For these purposes it is
convenient to put this equation into a Schr\"odinger-like form,
$-v_{nl}'' + U(r) \, v_{nl} = 0$, by rescaling $u_{nl} =
v_{nl}/(a^3 \, b)^{1/2}$, in which case the `potential' $U(r)$
takes the form
\eq
    U(r) = \frac{1}{a^2(r) b(r)} \left( \frac{n^2}{b(r)} - \lambda_{nl}
    \right) + \frac12
    \, \frac{d^2}{dr^2} \Bigl[ \ln(a^3 \, b) \Bigr] + \frac14 \,
    \left[ \frac{d}{dr} \ln (a^3 \, b) \right]^2 \, .
\eeq
A plot of this potential is given in figure \pref{fig:Uofr}. Near
$r = r_3$ we have $b(r) \approx b_3' (r - r_3)$ and $a(r) \approx
1$ and so in this region $U(r) \approx U_3 /(r - r_3)^{2}$ with
$U_3 = (n/b_3')^2 - \frac14$. We shall see later that positive
3-brane tension requires the defect angle $\veps_3 = 1 - b_3'/2$
to be positive, implying $b_3' < 2$. This in turn ensures that
$U_3 > 0$ for any nonzero $n$. For $r \to \infty$, on the other
hand, we have $a(r) = r/r_3$ and $b(r) \approx b_1$ and so $U(r)
\approx U_\infty /r^2$ with $U_\infty = \frac34 + (n^2/b_1 -
\lambda_{nl})(r_3^2/b_1)$. Clearly $U_\infty \ge 0$ for
$\lambda_{nl} \le (3b_1/4r_3^2) + (n^2/b_1)$.

We seek the eigenstates of this potential having zero energy, and
for these the region around $r = r_3$ is classically allowed
provided $U_3<0$. For $n \ne 0$ we see that both $U_3$ and
$U_\infty$ are positive for small enough $\lambda_{nl}$, and for
these choices there is typically no classically-allowed region for
which $U \le 0$. This shows that the least-massive $n \ne 0$ bulk
states have masses which are of order $m^2 \sim \lambda_{min} =
(3b_1/4 r_3^2) + n^2/b_1$, as expected.

The least massive bulk KK states must therefore have $n = 0$, in
which case $U_3 = -\frac14$ and $U_\infty = \frac34 - \lambda_{0l}
r_3^2/b_1$. If $\lambda_{0l} < 3b_1/4r_3^2$, then $U(r)$ is
negative near $r = r_3$ and positive at large $r$, implying $U$
must pass through zero at least once for finite $r$. Denoting the
smallest zero of $U$ by $r_z$, we see the existence of zero-energy
states localized near the 3-brane in the classically-allowed
region $r_3 < r < r_z$.

The eigenstates for this system in the WKB approximation are
\eq
    u_{nl}^\pm \approx \frac{A_{nl}^\pm}{\left[ a^3(r) \, b(r) \right]^{1/2}}
    \, \exp\left[\pm i \int_{r_3}^r dr' \,
    \sqrt{-U(r')} \right] \, ,
\eeq
where $A_{nl}^\pm$ are constants. These behave like $u^\pm_{nl}
\sim (r-r_3)^{\alpha_\pm}$ as $r \to r_3$, where $\alpha_\pm = -
\, \frac12 \pm i \sqrt{- U_3}$. These states are
therefore only marginally normalizable at $r = r_3$, using the
required norm: $2\pi \int_{r_3}^{r_4} a \, |u_{nl}^\pm|^2 \, dr$.
Since they are localized within $r < r_z$, there is a discrete
spectrum of eigenvalues for these lowest-energy $n=0$ states, and
the spacing of these eigenvalues should be independent of $r_4$ in
the limit $r_4 \gg r_z$. This indicates that the bulk KK modes are
generically independent of $r_4$ as $r_4$ is made large,
indicating that the KK gap remains fixed in this limit. If all
scales other than $r_4$ are chosen at the TeV scale, we therefore
expect the spectrum of massive bulk KK modes to also start in the
TeV region. In addition to these modes there will generally also
be a few bulk (and possibly 4-brane) massless modes (like the
graviton), which can appear in the low-energy, sub-TeV 4D
effective theory.

In summary, we have been led to a warped relatively-large
extra-dimensional scenario \cite{ADD}, with TeV physics on our
brane coupled to bulk modes which are generically at the TeV
scale, and to very weakly coupled physics at both TeV and
intermediate scales on the 4-brane. Although we have found that
the electroweak hierarchy requires the proper size of these extra
dimensions to be quite large compared to microscopic scales, the
large values we find for masses of the bulk KK modes makes this theory
safe with respect to tests of Newton's Gravitational Law
on submillimeter scales. The large KK masses also allow these
models to evade the serious astrophysical problems
\cite{SNProbs} which large-extra-dimensional models generically
have, and which are typically much worse within a supersymmetric
context \cite{susyADD}.

\subsubsection{Brane Boundary Conditions}
Clearly in these models an understanding of the hierarchy problem
involves an understanding of why $r_4$ should be so much larger
than $r_3$. Since this is determined by the brane properties, we
now turn to a more detailed description of how the branes couple.

Before plunging into the details, it is worth considering
the broader picture by first counting equations and
unknowns. There is a boundary condition at each brane for each
field in the problem. Given the symmetries of our solution this
gives rise to 3 conditions at the 3-brane (one each for the
dilaton, metric and Maxwell fields) plus 4 more at the 4-brane
(keeping in mind that the $(\mu\nu)$ and $(\theta\theta)$ metric
conditions on the 4-brane are independent). Thus there is a total of
7 conditions which must be solved for the various integration
constants of the solution.

Since there are 5 independent integration constants
($A$, $B$, $C$, $\alpha_3$ and $r_4$),  the system is
overconstrained and thus requires 2 independent conditions
on the
brane couplings of the model. We show in this section that these 2
constraints may be satisfied by choosing the dilaton coupling on
the 3-brane to vanish, plus a topological condition that relates
the coupling $g_2$ to $g$ (the coupling corresponding to the gauge
generator whose background field is nonzero).

With the above choices we therefore fix all of the
integration constants, showing that our ansatz has no moduli and
hence no classically massless dilaton or metric breathing
modes. In contrast, if the dilaton brane couplings are
chosen to preserve the classical scale invariance of the bulk
action, one combination of the integration
constants is a modulus which remains unfixed at the classical
level. The counting of constraints also changes, but leads to the
same conclusion as before. In this case the existence of the
undetermined modulus shows that the equations are not all
independent, so we must solve one fewer equation ({\it i.e.}\  6)
for one fewer ({\it i.e.}\  4) combination of parameters. This 
leaves the same two required adjustments among the coupling
constants as before.

Besides identifying how the couplings must be chosen in order to interpret
our solutions as being sourced by 3- and 4-branes, we also explicitly
solve for the dynamically-determined position of the 4-brane, $r_4$, and
in the process find what properties the branes must have in order to obtain a
large hierarchy $r_4 \gg r_3$.

\ssubsubsection{The 3 Brane}
We start with the 3-brane, for which
the dilaton and metric couplings in the brane action are taken to be
\eq
    S_3 = - T_3 \int d^4\xi \; e^{\lambda_3 \phi} \sqrt{- \det
    \gamma_{\mu\nu}} \, ,
\eeq
Here the induced metric is related to the 6D metric, $g_{MN}$,
and the 3-brane position, $x^M(\xi)$, by $\gamma_{\mu\nu} = g_{MN}
\, \partial_\mu x^M \partial_\nu x^N$. For coordinates $\xi^\mu =
x^\mu$, this becomes $\gamma_{\mu\nu} = g_{\mu\nu} + g_{mn}
\partial_\mu x^m \partial_\nu x^n$, where $\mu,\nu = 0,...,3$
and $m,n = 4,5$. For a brane at rest at $r=r_3$ we also have $x^m
= 0$. The quantities $T_3$ and $\lambda_3$ are the physical
3-brane properties which we wish to relate to the bulk geometry.

This action adds source terms to the dilaton and Einstein
equations, eqs.~\pref{E:Beom}. If the three brane is located at
position $x_3^m$, the source terms are of the form
\eqa
    \Box \phi + (\cdots) &=& \lambda_3 \, T_3 \, \frac{e^{\lambda_3
    \phi}}{e_2} \, \delta^2(x - x_3) \nn \\
    R_{MN} + (\cdots)_{MN} &=& T_3 \, \frac{e^{\lambda_3 \phi}}{e_2}
    \, \Bigl( g_{\mu\nu} \delta^\mu_M \delta^\nu_{n} - g_{MN}
    \Bigr) \, \delta^2(x - x_3) \, ,
\eeqa
where $e_2 = \sqrt{\, \det g_{mn}}$. These $\delta$-function
sources imply nontrivial boundary conditions for the bulk fields
at the brane position, as may be determined by integrating the
field equations over a small volume of infinitesimal proper radius
about the 3-brane position. Assuming the metric, dilaton and
Maxwell fields to be continuous at the brane position, we learn
how the dilaton derivative and the curvature behave there.

The dilaton derivative at the 3-brane position becomes:
\eq \label{dil3a}
    \lambda_3 T_3 \, e^{\lambda_3 \phi}
    \Bigr|_{r = r_3}    = \sqrt{ab} \, \phi' \Bigr|_{r = r_3}
\eeq
which should be read as a condition relating $\phi$ and $\phi'$ at
the brane position, given the known couplings $T_3$ and $\lambda_3$.
Since $\phi'$ is bounded as $r \to r_3$ in the solution of
interest, using $a(r_3) = 1$, $e^{\phi(r_3)} = \alpha_3$ and
$b(r_3) = 0$ shows that the right-hand side vanishes, and so
\eq \label{dil3b}
    \lambda_3 T_3 \alpha_3^{\lambda_3}
    =  0  \, .
\eeq
Since we do not wish to allow either $T_3$ or $\alpha_3$ to
vanish, we take this last condition to require $\lambda_3 = 0$.

A similar argument applied to the curvature singularity implies
the standard relation between the conical defect angle and the
3-brane tension \cite{Deser}:
\eq \label{met3}
    T_3 =  \Delta \theta = 2 \pi \left[1  -  \frac{b'(r_3)}{2}
    \right] = 2 \pi \, \left[1 - \frac{B}{r_3^3} \,
    \left( \frac{1 + A^2}{1 - A^2} \right) \right] \, .
\eeq
(recall that our units satisfy $\kappa_6^2 = 8\pi G_6 = 1$). We regard
this solution as fixing the value of $B$ once $T_3$ and $A$ are given.
Notice that if $T_3 > 0$ then we must require $b'(r_3) < 2$, and
so $A$ and $B$ must satisfy $|1 - A^2| \, r_3^3 > |B| (1 + A^2)$.

The 3-brane condition satisfied by the Maxwell field is found by
expressing the field strength in terms of a gauge potential as in
eq.~\pref{gaugepot}. The integration constant $C$ in this
expression is fixed by the requirement that there be vanishing
magnetic flux through an infinitesimal surface enclosing the
3-brane position. This is equivalent to demanding that the gauge
potential of eq.~\pref{gaugepot} vanish as $r \to r_3$, and so
\eq \label{gauge3}
    A^{\hat\alpha} = {A\, g_2 \alpha_3  r_3\over 2} \left( 1 -
    \frac{r_3^2}{r^2} \right) \, \exd \theta \,.
\eeq

In summary, the above equations determine how the 3-brane
properties $T_3$ and $\lambda_3$ are related to properties of the
bulk field configuration. In particular, the two integration
constants $B$ and $C$ are fixed by eqs~\pref{met3} and
\pref{gauge3}. On the other hand, the dilaton condition,
eq.~\pref{dil3a}, in general implies that the dilaton should be
singular at the 3-brane position, and so cannot be satisfied for
the smooth dilaton configuration considered here unless the
3-brane does not couple to the dilaton field at all --- {\it i.e.}\ 
$\lambda_3 = 0$. Both of the integration constants $\alpha_3$ and
$A$, as well as the 4-brane position $r_4$, then remain
undetermined by the 3-brane properties, and so are arbitrary at
this point. They are ultimately determined by the physics of the
4-brane, which is also what determines the volume of the internal
two dimensions.

\ssubsubsection{The 4 Brane}
We next ask what properties the bulk solution implies for the
4-brane which we assume terminates the extra dimensions at $r =
r_4 > r_3$. The precise nature of the conditions we obtain depends
on the kinds of 4-brane couplings we are prepared to entertain,
but in addition to the usual Nambu action it must also contain the
physics whose currents are generated by the electromagnetic fields
in the bulk. We start by considering the simplest
case, corresponding to the St\"uckelberg action for a
superconducting 4-brane
\eq \label{4braneaction}
    S_4 = - \int d^5\xi \; \sqrt{- \det
    \gamma} \, \Bigl[ T_4 \, e^{\lambda_4 \, \phi} + \sfrac12 \,
    e^{\zeta_4 \phi}
    \gamma^{ps} (\partial_p \,\sigma - q A_p) (\partial_s
    \,\sigma - q A_s)  \Bigr],
\eeq
where $\gamma_{pq}$ is the 4-brane's induced metric, $T_4$ is its
tension, $\lambda_4$ and $\zeta_4$ are dilaton couplings, and
$\sigma$ is a Goldstone mode living on the brane which arises due
to an assumed spontaneous breaking of the electromagnetic gauge
invariance. $q$ is a dimensionful quantity describing the energy
scale of this symmetry breaking. The derivative $D_p\, \sigma =
\partial_p \, \sigma - q A_p$ is gauge invariant given the
transformation rules $\delta A_p = \partial_p\, \omega$ and
$\delta \sigma = q\omega$.

With these choices the field equations for $\sigma$, $\phi$ and
$A_M$ at the position of the boundary 4-brane become
\eqa \label{4braneeqs}
    {\delta S_4 \over \delta \sigma} &=&
    \partial_p \Bigl[ e_5 \, e^{\zeta_4 \phi} D^p \sigma
    \Bigr]_{r=r_4} = 0 \nonumber \\
    \Bigl[ n_M \partial^M \phi \Bigr]_{r = r_4}
    &=& \frac{1}{e_5} \, {\delta S_4 \over \delta \phi}
    = - \Bigl[ \lambda_4 T_4 e^{\lambda_4 \phi} + \sfrac12
    \, \zeta_4 \, e^{\zeta_4 \phi} (D\sigma)^2 \Bigr]_{r=r_4} \\
    \left[ n_M \, e^{-\phi}\, F^{Mp} \right]_{r=r_4}
    &=& \frac{1}{e_5} \, {\delta S_4 \over \delta A_p} =
     q \, e^{\zeta_4 \phi(r_4)} \, D^p \sigma  \, , \nonumber
\eeqa
where $n_M \, dx^M = dr/\sqrt{g^{rr}}$ is the outward-pointing
unit normal to the surface $r = r_4$, and $e_5 = \sqrt{-\det
\gamma}$ is the volume element on the 4-brane.
In what follows we further simplify these expressions by choosing
the gauge $\sigma = 0$ on the 4-brane, in which case $D_p \,
\sigma = - q A_p$.  Using \pref{gaugepot}, we see that the first
condition of \pref{4braneeqs} is trivially satisfied since
$A_\theta$ is independent of $\theta$.

The last of these equations is the electromagnetic boundary
condition at the 4-brane position which expresses how much surface
current must flow in order to maintain the given magnetic flux,
$A$. We find
\eq \label{max4a}
     n_M g^{MN} \left. e^{-\phi} \, F_{Np}
     \, \right|_{r = r_4} = \sqrt{g^{rr}} \left. e^{-\phi}
     \, F_{rp} \, \right|_{r = r_4}
     = - q^2 \,
     \, e^{\zeta_4 \phi} \, A_p\, \Bigr|_{r = r_4} \, , \eeq
which takes the more explicit form
\eq \label{max4b}
    g_2 \, A \sqrt{[a b]}_{r_4} \,\left( \frac{r_3}{r_4} \right)^2
     = - \, q^2 \,
    \left( \alpha_3 \, { r_3\over r_4}\right)^{\zeta_4} \,
    A_\theta (r_4) \, .
\eeq

We regard this as an equation for $C_4$, where the gauge potential
at $r = r_4$ is written near the 4-brane as $A_\theta(r) =
\left[C_4 - \frac12 \, A \, g_2 \alpha_3 r_3^3/r^2 \right]$,
leading to
\eq
    C_4 = \left( \frac{A \, g_2 r_3^2}{r_4^2} \right) \left[
    \frac{\alpha_3 r_3}{2} - \frac{\sqrt{[ab]}_{r_4}}{q^2}
    \,
    \left( \alpha_3 \, { r_3\over r_4}\right)^{-\zeta_4}
    \right] \,.
\eeq
This can only differ from our 3-brane determination, $C_3 =
\frac12 \, A \, g_2 \alpha_3 r_3$, by at most a periodic gauge
transformation: $C_3 - C_4 = N/g$, where $N$ is an integer and $g$
is the gauge coupling for the gauge generator $\hat\alpha$ whose
background field is turned on (and so $g = g_2$ if $\hat\alpha \in
SU(2)$). This implies the constraint
\eq \label{fluxquantization}
    \frac{A \, g_2 \alpha_3 r_3 }{2} \left(1 - \frac{r_3^2}{r_4^2}
    \right) + \frac{\sqrt{[ab]}_{r_4}}{q^2} \,
    \left( \frac{A \, g_2 r_3^2}{r_4^2} \right)
    \left( \alpha_3 \, { r_3\over r_4}\right)^{-\zeta_4}
    = \frac{N}{g} \, .
\eeq
which is a flux-quantization condition,
restricting $A$ to take discrete values, labeled by an
integer $N$.

The second equation of \pref{4braneeqs}, the dilaton equation of
motion near the 4-brane, determines the radial derivative of
$\phi$ at $r = r_4$.
\eq
    n_M g^{MN} \partial_N \phi \Bigr|_{r = r_4} =
    \sqrt{g^{rr}}\, \partial_r \phi \Bigr|_{r = r_4} = -
    \Bigl[ \lambda_4 T_4 e^{\lambda_4 \phi} + \sfrac12 \, q^2
    \, \zeta_4 e^{\zeta_4 \phi} g^{\theta\theta} A_\theta^2 \Bigr]_{r = r_4}
    \,.
\eeq
{}From this equation it follows that the 4-brane dilaton boundary
condition is
\eq \label{dil4}
       - \sqrt{ab} \, \phi' \Bigr|_{r = r_4} = \left. \frac{\sqrt{ab}}{r}
       \right|_{r=r_4} =
    \left[
     \lambda_4 T_4 \, e^{\lambda_4 \phi} + { q^2 \zeta_4 e^{\zeta_4 \phi}
     A_\theta^2 \over 2\, a b} \right]_{r =
    r_4}  \, .
\eeq

Finally we consider the junction conditions for the metric elements
at the 4-brane, which express the response of the bulk metric to the
brane tension. The boundary couplings to the metric are obtained
from the 4-brane action, plus the Gibbons-Hawking
extrinsic-curvature term \cite{GHterm} (which is proportional to
integral over the boundary of the trace of the boundary's
extrinsic curvature). The result relates the extrinsic curvature
$K_{pq}$ of the 4-brane to the 4-brane stress energy, $S_{pq}$,
according to
\eq
    S_{pq} = - K_{pq} + g_{pq} g^{rs} K_{rs} \,.
\eeq

The brane stress energy to be used in this condition is
\eq \label{4branestress}
    S_{pq} = - \Bigl[T_4 e^{\lambda_4 \phi} + \sfrac12 \,
    e^{\zeta_4 \phi} (D\sigma)^2 \Bigr] g_{pq} + e^{\zeta_4\phi}
    D_p\, \sigma D_q \, \sigma \, .
\eeq
The extrinsic curvature of the surface $r = r_4$ is given by
\eq
    K_{pq} = - \left( \Gamma^M_{pq} \, \hat{n}_M
    \right)_{r = r_4} = -\left( \frac12 \, \sqrt{\, g^{rr}} \,
    g'_{pq} \right)_{r = r_4} =
    \pmatrix{K_{\mu\nu} & 0 \cr
    0 & K_{\theta \theta} \cr} \, ,
\eeq
where $\hat{n} = \hat{n}_M \, \exd x^M = -\exd r/\sqrt{\, g^{rr}}$
is the unit normal pointing into the bulk. The indices $p,q =
0,...,4$ include the four maximally-symmetric coordinates,
$\mu,\nu = 0,...,3$, and $\theta$. The components of $K_{pq}$ so
obtained are
\eq \label{extrinsicK}
    K_{\mu\nu} =  -\, \frac{\sqrt{ab}}{2}\left(\frac{a'}{a}\right)
      \, g_{\mu\nu} \qquad
    \hbox{and} \qquad K_{\theta\theta} = -\, \frac{\sqrt{ab}}{2} \left( \frac{a'}{a}
   + \frac{b'}{b} \right)
    g_{\theta\theta} \, ,
\eeq
implying that the combination which appears in the jump
conditions, ${\cal K}_{pq} = K_{pq} - g_{pq} g^{rs}K_{rs}$, is
given by
\eq \label{calKresult}
    {\cal K}_{\mu\nu} =  \frac{\sqrt{a b}}{2} \, \left( \frac{4 a'}{a}
    + \frac{b'}{b} \right)  \, g_{\mu\nu} \qquad \hbox{and} \qquad
    {\cal K}_{\theta\theta} =  \frac{\sqrt{a b}}{2} \,
    \left(\frac{4 a'}{a} \right) \,
    g_{\theta\theta}  \, .
\eeq

Using this extrinsic curvature with the stress-energy of
eq.~\pref{4branestress} implies the two conditions
\eqa \label{met4a}
    \Bigl[ T_4 \, e^{\lambda_4 \phi}  +  \sfrac12 \, q^2
    e^{\zeta_4 \phi} \, g^{\theta\theta} A_\theta^2  \Bigr]_{r = r_4}
    &=& \frac{\sqrt{a b}}{2} \left( \frac{4 a'}{a} + \frac{b'}{b} \right)_{r =
    r_4} \nonumber \\
    \hbox{and} \qquad q^2 e^{\zeta_4\phi} A_\theta^2 \, g^{\theta\theta}
    \Bigr|_{r = r_4}
    &=& \frac{\sqrt{ab}}{2} \left(
    \frac{b'}{b} \right)_{r = r_4} \, .
    \eeqa
Two conditions arise in this case because the electromagnetic
currents which support the bulk magnetic field generate an
asymmetric stress in the $\theta$ direction, ensuring that the stress
energy is not proportional to the metric, $g_{pq}$. We may use one
of these two conditions to determine the one remaining
undetermined integration constant, $r_4$. This leaves the other as
a redundant condition, which in general has a solution only for
specific choices for the couplings $T_3$ or $\Q$. (The latter coupling
will be introduced in section 5.2.)

Notice also that the second of eqs.~\pref{met4a} only has a
solution if $b'(r_4) > 0$, and inspection of the figure shows that
this is only possible if either: ($i$) $B > 0$ (and hence $|A| <
1$); or ($ii$) $B < 0$ (implying $|A| > 1$) and $r_4 < r_*$, where
$r_*$ is the position at which $b'(r_*)=0$.

\FIGURE{ \centerline{\epsfxsize=4.in\epsfbox{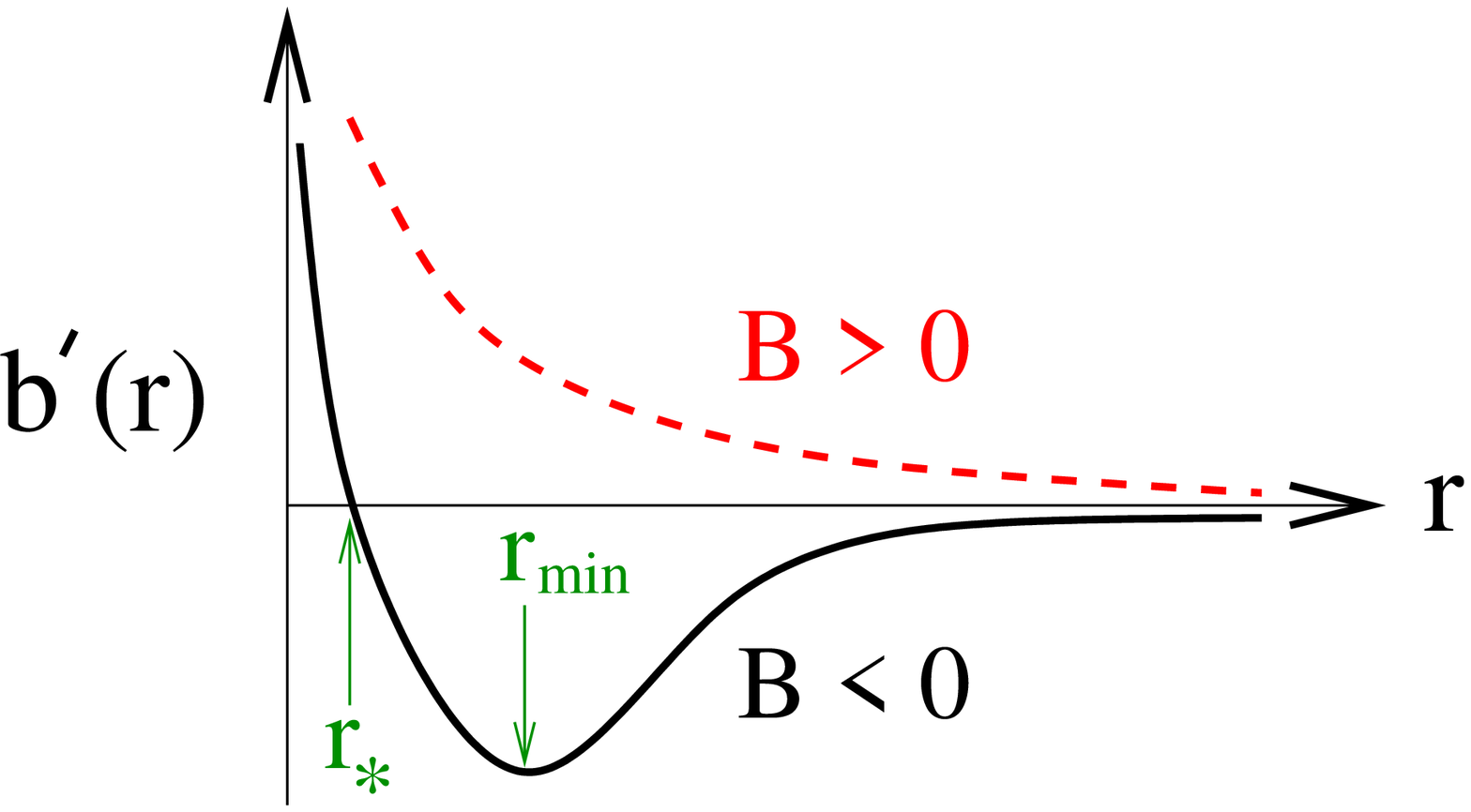}}
\caption{\small $b'(r)$ vs.\ $r$ in the two cases $B<0$ and $B>0$.
} }

We now turn to a more explicit solution of these last three
4-brane boundary equations ({\it i.e.}\ dilaton and metric
conditions) to see what brane properties are required in order to
obtain a large hierarchy, $r_4 \gg r_3$.

\ssubsubsection{Conditions for a Solution}
Our goal now is to solve eqs.~\pref{dil4} and \pref{met4a} to
determine $r_4$, $\alpha_3$ and $A$. We treat these three
variables as independent, although in so doing we redundantly
determine $A$, which must also satisfy
eq.~\pref{fluxquantization}. It is the reconciliation of these two
conditions which requires us to adjust the coupling $g$ of
eq.~\pref{fluxquantization} to $g_2$.

To solve these conditions it is useful to define the following
auxiliary quantities,
\eq
    D = \Bigl[ \frac{q^2}{2} \,  e^{\zeta_4 \phi} g^{\theta\theta}
    A_\theta^2 \Bigr]_{r_4},
    \qquad
    E = \Bigl[ T_4 e^{\lambda_4 \phi} \Bigr]_{r_4},
\eeq
and first solve for $E$, $D$ and $r_4$. The relevant three 4-brane
conditions can be written as
\eqa \label{EDeqn}
    E + D &=& \frac{\sqrt{[ab]}_{r_4}}{2} \left( {4a'\over a} +
    {b'\over b}\right)_{r_4}\\
    \label{Deqn}
    D &=& \frac{\sqrt{[ab]}_{r_4}}{2} \, \left( {b'\over 2b}
    \right)_{r_4} \\
    \label{ED2eqn}
    \lambda_4 E + \zeta_4 D &=& \frac{\sqrt{[ab]}_{r_4}}{r_4}
\eeqa

Let us first concentrate on $r_4$, by eliminating $E$ and $D$ from
these equations, leading to the result $4(1- 2\lambda_4) \, b(r_4)
= (\lambda_4 + \zeta_4) \,r_4 \, b'(r_4)$, or:
\eq
    \frac{2 b_2 [\zeta_4 - \lambda_4 + 1]}{r_4^4} +
    \frac{B [\zeta_4 - 3\lambda_4 + 2]}{r_4^2} -
    2(1-2\lambda_4) \, b_1 = 0 \, .
\eeq
This may be solved to give
\eq \label{r4result}
    \frac{2 b_2}{r_4^2} = \frac{ |B|}{2\, (\zeta_4 - \lambda_4 +
    1)}
    \left[ (3 \lambda_4 - \zeta_4 -2) \eta_B \pm \Delta \right] \, ,
\eeq
with
\eq
    \Delta = \left[ (3 \lambda_4 - \zeta_4 - 2)^2 + \frac{16 A^2
    (\zeta_4 - \lambda_4 + 1)(1 - 2\lambda_4)}{(1 - A^2)^2}
    \right]^{1/2} \, ,
\eeq
where $\eta_B = B/|B|$ and the sign in front of $\Delta$ in
eq.~\pref{r4result} is the choice which makes the overall result
positive (and so is equal to the sign of the product $(\zeta_4 -
\lambda_4 + 1)(1 - 2\lambda_4)$). We also use here the result
$b_1b_2 = B^2 A^2/(1 - A^2)^2$.

Since the hierarchy is determined by the ratio $r_3/r_4$, it is
useful to divide eq.~\pref{r4result} by the earlier result,
eq.~\pref{rthree}, in the form $2b_2/r_3^2 = -B + \sqrt{B^2 + 4
b_1b_2} = 2B A^2/(1-A^2)$. This gives
\eq \label{hiercond}
    \left( \frac{r_3}{r_4} \right)^2 =
    \frac{|1 - A^2|}{4 A^2 } \left[ \frac{(3 \lambda_4 - \zeta_4
    -2)\eta_B
     \pm \Delta}{(\zeta_4 - \lambda_4 +
    1)} \right] \, .
\eeq

Several conclusions may be drawn from these expressions.

\begin{itemize}
\item Eq.~\pref{hiercond} clearly shows that the hierarchy is completely
determined by the magnetic flux, $A$, and the 4-brane dilaton
couplings, $\lambda_4$ and $\zeta_4$.
\item In the limit $A = 1 - \epsilon$ with $|\epsilon| \ll 1$ we
have $\Delta \approx (2/|\epsilon|) \Bigl[ (\zeta_4 - \lambda_4 +
1) (1 - 2 \lambda_4) \Bigr]^{1/2}$ and so
\eq
    \left( \frac{r_3}{r_4} \right)^2 \approx \sqrt{ \frac{1-
    2\lambda_4 }{\zeta_4 - \lambda_4 + 1}} \,.
\eeq
This limit only makes sense (for real $r_4$) if $1-2\lambda_4$ and
$\zeta_4 - \lambda_4 + 1$ share the same sign. A hierarchy is in
this case ensured if $\lambda_4$ is chosen close to $\frac12$ but
with $\zeta_4$ not close to $-\,\frac12$.
\item If $|A| \ll 1$, then $B > 0$ and real solutions for $r_4$
exist provided $(\zeta_4 - 3 \lambda_4 + 2)$ and $(1 -
2\lambda_4)$ share the same sign. In this case
\eq
    \Delta \approx |3 \lambda_4 - \zeta_4 - 2| + \frac{8 A^2
    (\zeta_4 - \lambda_4 +1)(1 - 2\lambda_4)}{|3\lambda_4 -
    \zeta_4 - 2|} \, ,
\eeq
and so the hierarchy is found to be
\eq
     \left( \frac{r_3}{r_4} \right)^2 \approx \frac{2
     (1-2\lambda_4)}{\zeta_4 - 3 \lambda_4 + 2} \, .
\eeq
Again a large hierarchy is obtained if $\lambda_4$ is adjusted to
be close to $\frac12$, keeping $\zeta_4$ not too close to $- \,
\frac12$.
\item In the special case $\lambda_4 = - \zeta_4 = \frac12$, the
radius $r_4$ remains completely undetermined by the metric/dilaton
conditions for any $A$. This acts as a check on our calculations,
as we shall see in subsequent sections that in this limit the
4-brane action preserves the classical bulk scale invariance. In
this case one combination of integration constants cannot be
determined from the boundary conditions, and we expect to have a
massless modulus in the 4D spectrum. If the modulus
is taken to be $r_4$, then a large hierarchy may always be chosen
by moving along this flat direction out to large values of
$r_4/r_3$.
\item If $\lambda_4 = \frac12$ but $\zeta_4$ is kept general, then
there is no solution. This can be seen because there are
contradictory conditions for $A$. On one hand, the relation fixing
$r_4$ reduces in this case to $b'(r_4) = 0$, which is only
possible if $B < 0$ and so $|A| > 1$, since $r_4^2 = r_*^2 =
-2b_2/B$. On the other hand, the condition $b'(r_4) =
0$ in the metric matching conditions implies $D = 0$ and so also
$q A_\theta(r_4) = 0$. Consequently the 4-brane stress
energy is $SO(4,1)$ invariant, and must therefore be pure tension. In this
limit the 4-brane electromagnetic boundary condition,
eq.~\pref{max4b}, implies $A = 0$, contradicting the
earlier condition $|A|>1$.
\item Similarly, if $\zeta_4 = -\frac12$ with $\lambda_4$ kept
general, then $r_4$ becomes fixed by the condition $b(r_4) = 0$.
We discard this degenerate case since it corresponds to a bulk with one
less dimension, where $r_4 = r_3$.
\end{itemize}

In the generic case, with $r_4$ determined, we can solve
eqs.~\pref{EDeqn} and \pref{Deqn} for $E$ and $D$. $E$ immediately
determines the value of the dilaton at the 4-brane, which can be
taken to $\alpha_3$. As stated above, eq.~\pref{Deqn} then
provides a second determination of $A_\theta(r_4)$, and so also
$A$. The result obtained in general need not be consistent with
eq.~\pref{fluxquantization}, and so requires an adjustment of the
coupling constant $g$ relative to $g_2$. Alternatively, we can
adjust the 4-brane symmetry-breaking scale, $q$.

The scale-invariant case ($\lambda_4 = \frac12$ and $\zeta_4 =
-\frac12$) is similar. Here we may solve eqs.~\pref{EDeqn},
\pref{Deqn} and \pref{max4a} for $x = r_3/r_4$ and $A$ in terms of
$g_2$, $T_4$ and $q$. We then use the 3-brane tension condition to
fix $B/r_3^3$ and this, with the definition of $r_3$ ({\it i.e.}\ 
eq.~\pref{rthree2}), gives $\alpha_3 r_3$ purely in terms of
couplings and tensions. Since the flux-quantization condition
itself is a function only of the combinations $\alpha_3 r_3$ and
$B/r_3^3$, it provides a redundant constraint whose satisfaction
requires an adjustment of $g$ or $q$. We find that a large hierarchy,
$r_3 \ll r_4$, may be obtained in this case, for example by choosing
$q$ to be small. The details of this solution are provided in an
appendix (section 7).

In summary, we see that our solution in general only
describes the back reaction of the bulk fields to a 3- and 4-brane
for specific choices of brane coupling, $\lambda_3 = 0$, and
subject to a constraint which relates $g$ and $g_2$.

\section{Salam-Sezgin Supergravity in 6D}
In this section we present a warped compactification which is a
solution of the Salam-Sezgin chiral six-dimensional
supergravity-supermatter system. We begin by recapitulating the
relevant features of this model \cite{MS,NS,SS}.

\subsection{The Model}
The field content of Salam-Sezgin supergravity consists of a
supergravity-tensor multiplet consisting of a metric ($g_{MN}$),
antisymmetric Kalb-Ramond field ($B_{MN}$, with field strength
$G_{MNP}$), dilaton ($\phi$), gravitino ($\psi_M^i$) and
dilatino ($\chi^i$). The fermions are all real Weyl spinors, 
satisfying $\Gamma_7 \psi_M = \psi_M$ and $\Gamma_7 \chi = - \chi$
and so the model is anomalous unless it is coupled to an
appropriate matter content \cite{AGW}. The appropriate chiral 6D
matter consists of a combination of gauge multiplets,
containing gauge potentials ($A_M$) and gauginos ($\lambda^i$),
and $n_{\sss H}$ hyper-multiplets, with scalars $\Phi^a$ and
fermions $\Psi^{\hat{a}}$. The index $i = 1,2$ is an $Sp(1)$
index, $\hat{a} = 1,\dots,2n_{\sss H}$ and $a = 1,\dots,4n_{\sss
H}$. The gauge multiplets transform in the adjoint representation
of a gauge group, $G$. The $Sp(1)$ symmetry is broken explicitly
to a $U(1)$ subgroup, which is gauged.

The matter fermions are also chiral, $\Gamma_7 \lambda = \lambda$
and $\Gamma_7 \Psi^{\hat{a}} = - \Psi^{\hat{a}}$, but the
anomalies can be cancelled {\it via} the Green-Schwarz mechanism
\cite{GSAC}, for specific gauge groups and hypermultiplets
\cite{RSSS,6DAC}. An explicit example \cite{RSSS} of an
anomaly-free choice is $G = E_6 \times E_7 \times U(1)$, with the
hyper-multiplet scalars living on the noncompact quaternionic
K\"ahler manifold ${\cal M} = Sp(456,1)/(Sp(456)\times Sp(1))$.

The bosonic part of the classical 6D supergravity action is:
\eqa \label{E:Baction}
    e^{-1} {\cal L}_B &=& -\, \frac{1}{2 } \, R - \frac{1}{2 } \,
    \partial_{M} \phi \, \partial^M\phi  - \frac12 \, G_{ab}(\Phi) \,
    D_M \Phi^a \, D^M \Phi^b \cr
    && \qquad - \, \frac{1}{12}\, e^{-2\phi} \;
    G_{MNP}G^{MNP} - \, \frac{1}{4} \, e^{-\phi}
    \; F^\alpha_{MN}F_\alpha^{MN}
    -  e^\phi \, v(\Phi) \, .
\eeqa
Here the index $\alpha = 1, \dots, \hbox{dim}(G)$ runs over the
gauge-group generators, $G_{ab}(\Phi)$ is the metric on ${\cal M}$
and $D_m$ are gauge and K\"ahler covariant derivatives whose
details are not important for our purposes. We only require the
dependence on $\phi$ of the scalar potential for $\Phi^a = 0$,
which is $V(\phi,\Phi) = 2 \, g_1^2 \, e^{\phi}$. The
coupling $g_1$ denotes the $U(1)$ gauge coupling.

When the hypermultiplets and all but one of the gauge multiplets
are set to zero then the supersymmetry transformations reduce to
\eqa \label{E:susy}
    \delta e^{A}_{M} &=& {1 \over \sqrt{2}} \; \Bigl(
    \bar{\epsilon}\Gamma^{A}\psi_{M} - \bar{\psi}_{M}\Gamma^{A}\epsilon
    \Bigr) \nn\\
    \delta\phi &=& - \; {1\over \sqrt2 }
    \Bigl( \bar{\epsilon}\chi + \bar{\chi}\epsilon \Bigr) \nn \\
    \delta B_{MN} &=& \sqrt{2}  \, A_{[M}\delta A_{N]}
    + \frac{e^\phi}{2} \,
    \Bigl( \bar{\epsilon}\Gamma_{M}\psi_{N}-\bar{\psi}_{N}\Gamma_{M}\epsilon \nn \\
    && \qquad - \bar{\epsilon}\Gamma_{N}\psi_{M}
    +\bar{\psi}_{M}\Gamma_{N}\epsilon -
    \bar{\epsilon}\Gamma_{MN}\chi +
    \bar{\chi}\Gamma_{MN}\epsilon\Bigr)\\
    \delta\chi &=&
    \frac{1}{\sqrt2}\; \partial_{M}\phi \; \Gamma^{M}\epsilon
    + \frac{e^{-\phi}}{12} G_{MNP} \; \Gamma^{MNP}\epsilon \nn\\
    \delta\psi_{M} &=& \sqrt{2} \; D_{M}\epsilon +
    \frac{e^{-\phi}}{24} \; G_{PQR} \; \Gamma^{PQR}\Gamma_{M}\epsilon \nn\\
    \delta A_{M} &=&
    \frac{1}{\sqrt{2}} \Bigl(\bar{\epsilon}\Gamma_{M}\lambda
    - \bar{\lambda}\Gamma_{M}\epsilon\Bigr) e^{\phi/2} \nn\\
    \delta\lambda &=&
    \frac{e^{-\phi/2}}{4} \; F_{MN} \; \Gamma^{MN}\epsilon
    - \frac{i}{\sqrt{2}} \; g_1 \, e^{\phi/2} \, \epsilon \,
    , \nn
\eeqa
where the supersymmetry parameter is complex and Weyl: $\Gamma_7
\epsilon = \epsilon$.

\subsection{Compactification}
For our purposes we may set all gauge fields to zero except for a
single gauge potential, $A$, and we also set $\Phi^a = 0$. In this
section we derive a warped brane-world solution by continuing a
related nontrivial solution for the same system which was found
in ref.~\cite{FatPaper}. The solution in \cite{FatPaper} is given
by
\begin{eqnarray}
    && d s_{6}^{2} = -h(\rho)\, \exd \tau^{2} +\frac{\rho^{2}}{h(\rho)} \exd \rho^{2}
    + \rho^{2}
    \exd x_{0,4}^{2} \,, \nn\\
    && \phi(\rho) = - 2\, \ln \rho \,,\\
    && F_{\tau\rho} =\frac{\hat {\cal  A}}{\rho^5}\, \epsilon_{\tau\rho} \,, \nn
\end{eqnarray}
where $dx_{0,4}$ denotes a flat 4-dimensional spatial slice, and
%
\begin{equation}
h(\rho) = -\frac{2 {\cal M}}{\rho^{2}}  -
\frac{g_1^{2}\,\rho^{2}}{4} + \frac{{\hat {\cal
A}}^{2}}{16\,\rho^6} \,.
\end{equation}
This function has only a single zero for real positive $\rho$, and
${\cal M}$ and ${\hat {\cal A}}$ are integration constants which
can be positive or negative. This is not a brane-world solution
since the point where $h$ vanishes corresponds to a null Cauchy
horizon of the geometry.

A warped brane-world solution may be obtained from this one by
performing a suitable analytic continuation, in which we first
redefine the coordinate $r = \sfrac12 \, \rho^2$ so that the
previous solution takes the form
\begin{eqnarray}
    && d s_{6}^{2} = -h(r)\, \exd \tau^{2} +\frac{\exd r^{2}}{h(r)}
    + 2\,r [\exd x_1^2 + \exd x_{0,3}^{2}] \,, \nn\\
    && \phi(r) = - \ln (2 r)   \,,\\
    && F_{\tau r} =\frac{{ \hat{\cal A}}}{8\,r^3} \, \epsilon_{\tau r} \,, \nn
\end{eqnarray}
with
\begin{equation}
h(r)= \frac{2M}{r} - \frac{g_1^{2}\,r}{2}
                      + \frac{{\hat{\cal A}}^{2}}{128\,r^3} \,.
\end{equation}
Here we redefine the integration constant according to ${\cal M} =
-2M$, in anticipation of our later choice ${\cal M} < 0$. The new
solution is obtained by performing the analytic continuation
\begin{equation}
\tau \to i\, \theta\,, \qquad \qquad x_1 \to i t \qquad \qquad
                       \frac{\hat{\cal A}}{8} \to i {{\cal  A}}\,,
\end{equation}
in which case the it becomes:
\begin{eqnarray}
\label{newsol}
    && d s_{6}^{2} = 2\,r[-\exd t^2 + \exd x_{3}^{2}]
            +h(r)\, \exd \theta^{2} +\frac{\exd r^{2}}{h(r)} \,, \nn\\
    && \phi(r) =  - \ln (2r)  \,,\\
    && F_{\theta r} =-\frac{{\cal A}}{\,r^3} \, \epsilon_{\theta r} \,, \nn
\end{eqnarray}
with
\begin{equation}
h(r)= \frac{2M}{r} - \frac{g_1^{2}\,r}{2}
                      -\, \frac{{\cal A}^{2}}{2\, r^3} \,.
\end{equation}
This is the desired solution whose properties we now explore.

\subsubsection{Singularities and Supersymmetry}
Eq.\ \pref{newsol} describes a Lorentzian-signature solution provided $h(r) >
0$, and so it is useful to enumerate the zeroes of $h(r)$, which
occur at
\begin{equation}
    r^2_\pm = \frac{2\,M}{g_1^2} \left[ 1 \pm \sqrt{1 -
    \left( \frac{g_1\, {\cal A}}{2 M} \right)^2 } \right] \,.
\end{equation}
Since $h(r) < 0$ when $r \to \infty$ and $r \to 0$, the regime of
interest for a brane-world solution is the interval $r_- < r <
r_+$. This interval is not empty provided $M > \sfrac{1}{2} |g_1
{\cal A}| > 0$, a condition which we henceforth assume.

The geometry pinches off at the points $r = r_\pm$, at each of
which it generically has conical singularities. We therefore place
a 3-brane at each of these points when constructing a brane-world
model. Repeating the discussion of the previous sections shows
that the conical defect at $r = r_\pm$ is given by
\eq
    \varepsilon_\pm = 1 - \frac{|h'(r_\pm)|}{2} = 1 -
    \frac{g_1^2}{2 \, r_\pm^2} \left(r_+^2-r_-^2\right)\,
    .
\eeq
This last equality is obtained by writing $h(r) = -\frac12 \,
(g_1^2/r^3)(r^2 - r_+^2)(r^2 - r_-^2)$.

These conditions show that the defect angles are completely
determined by the two quantities $r_-/r_+$ and $g_1$. In
particular, one of the conical defects can be smoothed over if
$r_-/r_+$ is chosen appropriately. We find
\eq
    \veps_+ = 0 \quad \implies \quad
    \frac{r_-^2}{r_+^2} = 1 - \frac{2}{g_1^2}
     \qquad \hbox{and} \qquad
    \veps_- = 0 \quad \implies \quad
    \frac{r_+^2}{r_-^2} = 1 +
    \frac{2}{g_1^2}\, .
\eeq
Notice that the condition for the removal of the singularity at
$r_+$ requires a large coupling $g_1 > \sqrt2$, and so is only of
doubtful validity in a perturbative calculation such as ours.

\ssubsubsection{Supersymmetry}
This solution generically breaks supersymmetry, as is most easily
seen by specializing the $\chi$ supersymmetry transformation to
it, with the result
\eq
    \delta\chi =
    \frac{1}{\sqrt2}\; \partial_{M}\phi \; \Gamma^{M}\epsilon
    \,.
\eeq
This clearly cannot vanish because $\partial_M \phi \ne 0$.

\subsection{Brane Worlds}
In this section we examine the properties of the brane-world
scenario constructed from the warped solution given above. In this
case the construction requires two 3-branes, respectively located
at the conical singularities $r = r_\pm$, allowing us to interpret
these singularities as the gravitational back-reaction due to the
presence of the branes.

\subsubsection{Electroweak Hierarchy}
In the present instance the warp factor is $w(r) = 2\,r$ and so
the expression for the effective 4D Planck mass becomes
\eq
    M_p^2 = 2 \pi \int_{r_-}^{r_+} dr \, w(r) = 2 \pi(r_+^2 -
    r_-^2 ) = \frac{8 \pi \, M}{g_1^2} \sqrt{1 - x^2} \, ,
\eeq
where $x = g_1 {\cal A}/(2 M)$. For comparison, the physical mass
of a particle localized on the 3-brane located at $r = r_\pm$ is
\eq
    m_\pm = \mu_\pm \sqrt{w(r_\pm)} \, ,
\eeq
where the particle action is assumed to be proportional to
$g^{\mu\nu} \partial_\mu \chi \partial_\nu \chi + \mu_\pm^2
\chi^2$.

The hierarchy between these scales is therefore
\eqa
    \frac{M_p^2}{m_\pm^2} &=& \frac{4 \pi M}{g_1^2 \mu_\pm^2
    r_\pm}  \, \sqrt{1 - x^2} = \frac{2 \pi \, {\cal A}}{g_1 \mu_\pm^2 r_\pm}
    \, \left( \frac{\sqrt{1 - x^2}}{x} \right) \, \nonumber \\
    \frac{m_+^2}{m_-^2} &=& \frac{\mu_+^2 \, r_+}{\mu_-^2 \,
    r_-} = \frac{\mu_+^2}{\mu_-^2} \, \left( \frac{1 +
    \sqrt{1 - x^2}}{1 - \sqrt{1 - x^2}} \right)^{1/2} \, ,
\eeqa
so a large hierarchy can be achieved, for example, if all
dimensionful quantities are the same order of magnitude, except,
say, $M$, which we take to be much larger. The hierarchy is then
controlled by $x \ll 1$, or $g_1 {\cal A} \ll 2 M$, and in this
case the previous formulae for $r_\pm$ reduce to
\eq
    r_+^2 \approx \frac{4M}{g_1^2} \qquad \hbox{and} \qquad
    r_-^2 \approx \frac{{\cal A}^2}{4 M} \, ,
\eeq
and so $r_-/r_+ \approx x/2$. Clearly this does not really provide
a satisfactory explanation for the electroweak hierarchy, since
the desired scales are simply inserted into the higher-dimensional
solution.

If the gauge coupling $e^{\phi(r_-)}$ is assumed small, then the
solution guarantees the gauge coupling to be even smaller at $r =
r_+$ by an amount $e^{\phi(r_+)}/e^{\phi(r_-)} = r_-/r_+$.

\subsubsection{Brane Boundary Conditions}
To understand what the previous choices for ${\cal A}$ and $M$ mean
physically it is necessary to connect these integration constants
to brane properties.

The counting of boundary conditions proceeds as follows. As
before, the smoothness of the dilaton field at the 3-brane
positions precludes these branes from directly coupling to the
dilaton. Because this is also the choice which preserves the bulk
scale invariance, the metric condition at each 3-brane only
involves the scale-invariant ratio ${\cal A}/M$, implying a
topological constraint which relates the two tensions to one
another. The Maxwell boundary conditions at each 3-brane then
lead to contradictory conditions on the gauge potentials, which
imply a final topological restriction, also involving only
the ratio ${\cal A}/M$.

We are therefore led in this case to three kinds of constraints.
Two of these (vanishing 3-brane/dilaton charge, and the Maxwell
flux-quantization condition) are similar to those found earlier
for Romans' supergravity. Flux-quantization can be satisfied by
adjusting the background gauge coupling, $g$, in terms of the
coupling, $g_1$, appearing in the scalar potential. The third
restriction, relating the 3-brane tensions, has no counterpart for
Romans' supergravity and arises in the Salam-Sezgin case because
of the compactness of the internal two dimensions. (This
constraint is the analog of the condition of equal tensions which
arises in the unwarped case \cite{branesphere}.) In summary, we
are led in this model to a picture which is very similar to what
was encountered elsewhere for the unwarped solutions to
Salam-Sezgin supergravity.

\ssubsubsection{Dilaton and Metric}
Following the reasoning of previous sections we see that 3-branes
having the actions
\eq \label{3brane0action}
    S_\pm = - T_\pm \int_{r_\pm} d^4\xi \, e^{\lambda_\pm \phi} \, \sqrt{
    - \det \gamma} \, ,
\eeq
implies the dilaton couplings must satisfy
\eq
    \lambda_\pm  = 0 \, ,
\eeq
in order for our assumed solution to describe correctly the fields
generated by branes having the assumed action. This condition
follows from the smoothness of the dilaton at the 3-brane
positions.

Similarly, the brane tensions are related to the corresponding
conical defect angles by the conditions
\eq \label{Tpmconditions}
    T_\pm = 2\pi \left[ 1 - \frac{|h'(r_\pm)|}{2} \right] = 2 \pi
    \left[ 1 - \frac{g_1^2}{2\, r_\pm^2} \left(r_+^2-r_-^2\right)\right] \, ,
\eeq
from which we see that positive tensions imply that the  radii
$r_\pm$ must satisfy $\left({r_+}/{r_-}\right)^2 < 1+2/g_1^2$, or
in terms of $x = g_1 {\cal A}/(2M)$: $x^2 > 1- (g_1^2+1)^{-2}$.
Notice that large $r_+$, $r_+ \gg r_-$, therefore clearly requires $g_1
\ll 1$.

These last two brane boundary conditions determine only one of the
two integration constants, $M$ and ${\cal A}$, (or equivalently of
$r_+$ and $r_-$) because they depend on the ratio $r_-/r_+$,
and so can only determine the combination $x = g_1 {\cal A}/(2M)$.
The fact that the two tensions are both determined by the single
variable $x$ implies the existence of a constraint relating these
tensions. Eliminating $r_-/r_+$ from eq.~\pref{Tpmconditions}
gives
\eq \label{Tconstraint}
    \frac{T_+ - T_-}{2 \pi} - \frac{2}{g_1^2} \left(1 -
    \frac{T_+}{2\pi} \right) \left( 1 - \frac{T_-}{2\pi} \right) =
    0 \, .
\eeq
This is the analogue of the condition that the two 3-brane
tensions be equal, which obtains for the unwarped 2-sphere
solution \cite{branesphere}.

\ssubsubsection{Gauge Fields}
A similar condition applies at the position of each brane, which
follows from the nature of the brane coupling to the background
Maxwell field. For the action of eq.~\pref{3brane0action}, the
brane carries no flux, and so the flux through a small patch of
infinitesimal radius $\epsilon$ about each brane position must
vanish in the limit $\epsilon \to 0$. This condition applied to
both branes leads to a topological constraint which the parameters
of our solution must satisfy.

To see this, notice that the gauge potential for the magnetic
field strength, $F = ({\cal A}/r^3) \,  \exd r \wedge \exd \theta$
may be written
\eq \label{Apatch}
    A = \left( c - \frac{{\cal A}}{2 r^2} \right) \exd \theta \,,
\eeq
where $c$ is an integration constant. The condition that $F$ not
contain delta-function contributions at $r = r_\pm$ requires $A$
to vanish at these two positions, and this imposes contradictory
constraints on $c$: $c = c_\pm = {\cal A}/(2 \, r_\pm^2)$.
Consequently $F$ can only be nonsingular at {\it both} $r = r_+$ and
$r=r_-$
if eq.~\pref{Apatch} holds separately for two overlapping patches,
$P_\pm$, each of which includes only one of $r_+$ or $r_-$.

Although the gauge potential can take different values ($A =
A_\pm$ distinguished by constants $c_\pm$) on each of these
patches, $A_+ - A_-$ must be a gauge transformation. Periodicity
of the coordinate $\theta$ on the overlap then requires $c_+ - c_-
= n/g$, where $g$ is the gauge coupling appropriate for the
background gauge field which has been turned on. Combined with the
expressions for $c_\pm$ we find the requirement
\eq \label{topcond}
     \frac{{\cal A}}{2} \left( \frac{1}{r_-^2} - \frac{1}{r_+^2} \right)
     = \frac{2 M}{{\cal A}}
     \, \sqrt{1 - \left(\frac{g_1 {\cal A}}{2 M} \right)^2}
     = \frac{n}{g} \, .
\eeq
For the case of large, $r_+ \gg r_-$, this condition simplifies to
$2 M/{\cal A} \approx  n/g$, and so $r_-/r_+ \approx g_1 {\cal
A}/(4 M) \approx g\, g_1/(2n) \ll 1$. Since the ratio $r_-/r_+$ is
already fixed given $T_+$ or $T_-$, we instead read
eq.~\pref{topcond} as a condition relating $g$ to $g_1$.

Since all of these conditions only fix the ratio ${\cal A}/M$ and
none separately determine ${\cal A}$ or $M$, the overall scale of
the extra dimensions (say, its volume) remains undetermined. As
described in detail in the next section, this is consistent with
the scale invariance of the bulk equations which is not broken by
the 3-brane. Consequently ${\cal A}$ parameterizes a flat
direction, for which we expect a classically massless modulus in
the low-energy 4D theory.  This behavior is in contrast to that of
nonsupersymmetric versions of this model, lacking the dilaton,
 where the volume of the
extra dimensions is automatically stabilized in the presence of
nonvanishing gauge flux \cite{football}.

\section{Self-Tuning in Six Dimensions}
The solutions we have found have flat 4D slices for all values of
the various integration constants they involve. On the other hand,
we have found that regarding these solutions as being sourced by
simple 3- or 4-branes requires nontrivial relations amongst the
couplings of the model. It is natural to then ask whether these
choices were also required in order to adjust the 4D cosmological
constant to vanish, or if they are choices which are only forced
on us by our inability to find the general solution corresponding
to the fields set up by a generic brane configuration.

In this section we partially address this question by identifying
the source of the vanishing of the 4D cosmological constant in as
much generality as possible. In particular, we avoid use of the
detailed properties of the solutions, to see which features are
important (and which are not) for ensuring flatness in 4D. By
generalizing the argument of ref.~\cite{branesphere} we show here
that 4D flatness turns on the classical scale invariance of the
bulk equations, and so hinges on whether the dilaton/brane
couplings are chosen in a scale-invariant way. By contrast
flatness does not appear to depend in an important way on the
various topological conditions we have found.

To show this, in this section we explicitly integrate out the bulk
massive KK modes, which at the classical level amounts to setting
the fermionic modes to zero and eliminating the bosonic modes from
the action using their classical equations of motion. We are
therefore interested in the value of the action when evaluated at
the solution to the classical equations of motion.

Before performing this integration for the specific models
described above, we first recast the argument in its most general
form which exposes the connection to the classical bulk scale
invariance. Our arguments show that this self tuning property need
not be specific to six dimensions, and may rather be generic to
higher-dimensional supergravities. Making the connection to scale
invariance also allows a more precise comparison of these models
with the general self-tuning formulation given in
ref.~\cite{CCReview} (for discussions on the 5D case see for instance
\cite{adks,peter}). Making this connection explicit makes it
possible to ask whether there are loopholes to the no-go arguments
that the general four-dimensional analysis suggests. We reserve
our remarks concerning possible loopholes for the discussion,
Section 5.

\subsection{General Arguments}
The classical self-tuning properties of the 6D (and
higher-dimensional) supergravity equations follows from the
classical scale invariance which they enjoy. In this section we
review the general argument as to how this ensures a self-tuning
of the cosmological argument.

Consider therefore the interactions of generic matter fields
$\phi$ and the metric $g_{MN}$ in an $n$-dimensional bulk, coupled
to various brane modes on a set of $(n-1)$-dimensional boundary
branes. Our later application is to a 6-dimensional bulk bounded
by 5-dimensional 4-branes. 3-branes may also be included, in which
case the boundary contribution consists of a small circle of
infinitesimal radius which surrounds the 3-branes. We take the
action for the theory to be
\def\args{{\phi, g_{MN}}, \cdots}
\def\shortargs{\phi,\di\phi,g_{MN}, \cdots}
\def\Scl{{\cal L}}
\eqa \label{eq:bd:tot:lag}
    S &=& \int_M d^n x \; \Scl_B( \args ) + \int_{\di
    M} d^{n-1}x \; \Scl_b(\args) \cr
    &=&  \int_M d^n x \; \Scl_B + \int_{\di M} d^{n-1}x \;\Bigl[
    \Scl_b^0(\args) + \widehat\Scl_b(\args)
    \Bigr]
\eeqa
where $\Scl_B$ is the bulk lagrangian density, and $\Scl_b$ is the
brane action, which the second line splits into two pieces,
$\Scl_b^0$ and $\widehat{\Scl}_b$. $\Scl_b^0$ here consists of the
boundary pieces (such as the Gibbons-Hawking extrinsic-curvature
term \cite{GHterm}) which are required by the bulk action, while
$\widehat\Scl_b$ denotes the explicit brane action (such as the
Nambu action used above).

Higher-dimensional supergravity theories typically have the
following rescaling property for constant c:
\eqa
    \Scl_B(\args) &= & e^{-\omega_B c} \Scl_B(\phi-c,
    e^{sc} g_{MN},\cdots) \cr
    \Scl_b^0(\phi,g_{MN},\cdots) &=& e^{-\omega_B c}
    \Scl_b^0(\phi-c,e^{sc} g_{MN},\cdots) \cr
    \widehat\Scl_b(\phi, g_{MN}, \cdots ) &=& e^{-\omega_b c}
    \widehat\Scl_b(\phi-c,e^{sc}g_{MN}, \cdots).
\eeqa
When regarded as low-energy vacua of string theory this invariance
can be traced to the dependence on the dilaton, for which the
classical scale invariance is manifest in the string frame. The
ellipses in these equations denote any other fields, some of which
may transform under the rescaling symmetry.

Now comes the main point. We rewrite the action,
\myref{eq:bd:tot:lag} using the above relations, and take the
derivative with respect to $c$, setting $c=0$ afterwards.  Since
$S$ is independent of $c$, we have
\eqa
    0= \dii{S}{c} &=& - \int_M d^nx \; \left[\omega_B
    \Scl_B( \args ) + \dii{\Scl_B}{\phi} \right] - \int_{\di M} d^{n-1}x \;
    \dii{\Scl_b}{\phi} \cr
     &-& \omega_B\int_{\di M} d^{n-1}x \;
    \Scl_b^0(g_{MN},\cdots)
    - \omega_b \int_{\di M} d^{n-1}x \; \widehat\Scl_b(\phi, g_{MN}, \cdots)
   \cr &+& \hbox{(terms
    vanishing on use of all but the $\phi$ equations of motion)} \,
    .
\eeqa
The key observation is that the terms involving differentiation with
respect to $\phi$ cancel when evaluated at the solutions to the
dilaton equation of motion. This cancellation arises between the
bulk and boundary terms, as may be seen from the bulk equations of
motion,
\eq
    \dii{\Scl_B}{\phi}
    = \di_M \left(\dii{\Scl_B}{\di_M\phi}\right) \equiv \di_M V^M  \,
    ,
\eeq
together with their counterparts on the boundary:
\eq
    \dii{\Scl_b}{\phi} =
    \di_\mu\left(\dii{\Scl_b}{\di_\mu\phi}\right) - n_M V^M \, .
\eeq
Here $n_M$ denotes the outward pointing unit normal on the
boundary brane, and the last term in this equation arises due to
an integration by parts in the bulk. Keeping in mind that total
derivatives may be dropped on the boundary, we are finally left
with
\eq \label{eq:condition}
    0 = \omega_B \left[ \int_M d^nx \; \Scl_B +
    \int_{\di M} d^{n-1}x \; \Scl_b^0 \right] + \omega_b \int_{\di M}
    d^{n-1}x \; \widehat\Scl_b \,.
\eeq
We now use \myref{eq:condition} to eliminate all pieces of the
bulk action in favor of the brane action to get
\eq
    S =
    \left(1-{\omega_b \over \omega_B} \right) \int_{\di M}
    d^{n-1}x \; \widehat
    \Scl_b \, .
\eeq

For example, for the dilaton gravity such as arises in
D-dimensional supergravity we have $s = 2 \omega_B/(D-2)$ and
$\omega_B = 2$. Consequently $s=1$ in $D=6$ and $s = \frac12$ for
$D = 10$. Furthermore, $\widehat \Scl_b = e^{\lambda \phi}
\sqrt{-g}$ implies $\omega_b = -\lambda + ds/2$ where $d=n-1=p+1$
is the dimension of the world-volume for a $p$-brane. Therefore,
in the 6D case of present interest we have, using $\omega_B = 2$
and $s=1$,
\eq
    1-{\omega_b \over \omega_B} = 1+\frac{\lambda}{2} - \frac{d}{4} \, .
\eeq
Applied to a 4-brane we have $d=5$, and so $S = 0$ when evaluated
at the classical equations if $\lambda_4 = 1/2$. For a 3-brane we
instead use  $d=4$ to get $S = 0$ if $\lambda_3 = 0$.

We expect from these arguments that scale invariance should ensure
a classical self-tuning of the 4D cosmological constant for
configurations built using $\lambda_3 = 0$ 3-branes and $\lambda_4
= \frac12$ 4-branes. A similar argument applied to the 4-brane
action for the $\sigma$ fields implies self-tuning should occur if
the dilaton coupling of eq.~\pref{4braneaction} satisfies
\eq
    \zeta_4 = -1 + \frac{d-4}{2} \,,
\eeq
which reduces to $\zeta_4 = -\, \frac12$ for a 4-brane ($d=5$).

We now investigate more explicitly how these general arguments
work for the two 6D supergravity models considered in previous
sections.

\subsection{Romans' Supergravity}
To explicitly see how self-tuning works for Romans' supergravity,
consider the following expression for the effective 4D
cosmological constant, obtained by evaluating the classical action
with the bulk Kaluza-Klein modes integrated out at tree level.
Since this is equivalent to their elimination using their
classical equations of motion, we have
\eqa \label{rhoeffR}
    \rho_{\rm eff} &=& \left. T_3 a^2(r_3)\, e^{\lambda_3 \phi} \right|_{r= r_3} +
    \left. T_4 \, a^2(r_4) \,\int_0^{2\pi} d \theta \, e^{\lambda_4 \phi}
    \sqrt{g_{\theta\theta}} \right|_{r = r_4} \\
    &&\qquad + \int_M d^2y \; e_2 \, a^2 \left[\hf R_6 + \hf
    (\partial \phi)^2
     + \frac{1}{12} \, e^{-2\zeta \phi} \, G^2
    + \frac14 \, e^{-\phi} \, \Bigl(F^2 + \cF^2 \Bigr) \right. \nn\\
    && \qquad\qquad  \left.  - \frac12 \, g_2^2 \,
    e^\phi  + \frac{1}{8\sqrt2} \, \epsilon^{MNPQRS}
    B_{MN} \Bigl(F^\alpha_{PQ} F_{\alpha RS} + \cF_{PQ} \cF_{RS}
    \Bigr) \right]_{cl} \, , \nn
\eeqa
where the subscript `$cl$' indicates the evaluation of the result
at the solution to the classical equations of motion. For
simplicity we choose $q = 0$ and so neglect to the
contributions to $\rho_{\rm eff}$ of the superconducting currents.

Here we adopt a procedure for which the branes are represented as
delta-function contributions to the bulk equations of motion, and
so if $\widehat M$ is the two-dimensional bulk having the 4-brane
as a boundary, $M$ denotes the two-dimensional bulk manifold
obtained by gluing two copies of $\widehat M$ together at the
4-brane position. For future purposes it is important to recognize
that whereas $\widehat M$ has a boundary, $M$ does not unless we
introduce boundaries by excising small circles about the positions
of any 3-branes. This choice is purely a matter of convenience,
and we have verified that our conclusions are unchanged if we
instead work directly with $\widehat M$, keeping an explicit
boundary at the position of the 4-brane.

Eliminating the metric using the Einstein equation \pref{E:Beom}
allows the 6D curvature scalar to be replaced by
\eqa
    R_6 &=& - (\partial \phi)^2 + \frac32 \, g_2^2 \, e^\phi - \frac14 \,
    e^{-\phi} \, \Bigl( F^2 + \cF^2 \Bigr) \nn\\
    &&\qquad - {2 \over e_2} \,
    \,T_3 \, e^{\lambda_3 \phi}  \, \delta^2(x - x_3)
    - \frac52 \, T_4 \, e^{\lambda_4 \phi} \frac{1}{\sqrt{g_{rr}}}
    \, \delta(r - r_4) \,  .
\eeqa
Substituting this into eq.~\pref{rhoeffR} we find
\eqa \label{RemoveGR}
    \rho_{\rm eff} &=& \left. - \frac14 \, T_4 \, a^2(r_4) \,\int_0^{2\pi}
    d \theta \, e^{\lambda_4 \phi}
    \sqrt{g_{\theta\theta}} \right|_{r = r_4} \nn\\
    && \qquad +  \int_M d^2y \; a^2\,  e_2 \, \left[  \frac{1}{12}
     \, e^{-2\zeta \phi} \, G^2 + \frac18 \,
    e^{-\phi} \Bigl( F^2 + \cF^2 \Bigr) + \frac14 \, g_2^2 \, e^\phi
    \right. \nn\\
    && \qquad \left.  + \frac{1}{8\sqrt2} \, \epsilon^{MNPQRS}
    B_{MN} \Bigl(F^\alpha_{PQ} F_{\alpha RS} + \cF_{PQ} \cF_{RS}
    \Bigr)
    \right]_{cl}  \, .
\eeqa
Notice that the 3-brane tension cancels, just as in
ref.~\cite{branesphere}.

Repeating this process, to integrate out the dilaton classically,
allows us to use
\eqa
    \frac12 \, g_2^2 \, e^\phi + \frac14 \, e^{-\phi} \,\Bigl( F^2
    + \cF^2 \Bigr) + \frac{\zeta}{6} \, e^{-2\zeta \phi} \, G^2
    &=& - \Box \phi + \lambda_3 \, T_3 \, e^{\lambda_3 \phi}
    \frac{1}{e_2} \, \delta^2(x - x_3) \nn\\
    && \quad +  \lambda_4
    \, T_4 \, e^{\lambda_4 \phi} \frac{1}{\sqrt{g_{rr}}} \, \delta(r - r_4)
    \, ,
\eeqa
and further simplify eq.~\pref{RemoveGR} to become
\eqa \label{RemovePhi}
    \rho_{\rm eff} &=& \frac12 \left. \left(\lambda_4 - \, \frac12 \right)
    \, a^2(r_4)\, T_4 \,\int_0^{2\pi}
    d \theta \, e^{\lambda_4 \phi}
    \sqrt{g_{\theta\theta}} \right|_{r = r_4}
    +\left.  \frac12 \, \lambda_3 \, a^2(r_3)\, T_3 \, e^{\lambda_3 \phi}
    \right|_{r = r_3} \nn\\
    && \qquad +  \int_M d^2y \; a^2\, e_2 \, \left[
    \frac{1}{12} (1- \zeta)
     \, e^{-2\zeta \phi} \, G^2 - \frac12 \, \Box \phi
    \right.  \nn\\
    && \qquad \left. + \frac{1}{8\sqrt2} \, \epsilon^{MNPQRS}
    B_{MN} \Bigl(F^\alpha_{PQ} F_{\alpha RS} + \cF_{PQ} \cF_{RS}
    \Bigr)
    \right]_{cl} \, .
\eeqa

If we now integrate out $B_{MN}$ using its equation of motion, we
may write
\eq
    \frac{1}{8\sqrt2} \, \epsilon^{MNPQRS}
    B_{MN} \Bigl(F^\alpha_{PQ} F_{\alpha RS} + \cF_{PQ} \cF_{RS}
    \Bigr) = \frac12 \,B_{MN}  D_P \Bigl( e^{-2 \zeta \phi} \, G^{PMN}
    \Bigr) \, ,
\eeq
leading to
\eqa \label{RemoveB}
    \rho_{\rm eff} &=& \frac12 \left. \left(\lambda_4 - \, \frac12 \right)
    \, a^2(r_4) \, T_4 \,\int_0^{2\pi}
    d \theta \, e^{\lambda_4 \phi}
    \sqrt{g_{\theta\theta}} \right|_{r = r_4}
    +\left.  \frac12 \, \lambda_3 \, a^2(r_3)\, T_3 \, e^{\lambda_3 \phi}
    \right|_{r = r_3} \\
    && \quad - \int_M d^2y \; a^2\, e_2 \, \left[
    \frac{1}{12} (1 + \zeta)
     \, e^{-2\zeta \phi} \, G^2 + \frac12 \, \Box \phi
    - \frac12 \, D_P \Bigl( e^{-2 \zeta \phi} \, B_{MN} \, G^{PMN}
    \Bigr)
    \right]_{cl} \nn\\
    &=& - \,  \frac{1}{12} (1 + \zeta)
    \int_M d^2y \; a^2\, e_2 \,  e^{-2\zeta \phi} \, G^2
    + a^2(r_4) \left[ \pi \left(\lambda_4 - \, \frac12 \right)
    \, T_4 \,  e^{\lambda_4 \phi} \sqrt{g_{\theta\theta}}
    \right]_{r=r_4}\nn\\
     && \qquad + a^2(r_3) \left[ \frac12 \, \lambda_3 \, T_3 \, e^{\lambda_3 \phi}
    - \pi \, e_2 n_P \Bigl( \partial^P \phi  -
    e^{-2 \zeta \phi} \, B_{MN} \, G^{PMN} \Bigr)
    \right]_{r = r_3}
    \, . \nn
\eeqa
Here the surface integral for the 3-brane is evaluated on an
infinitesimal 5 dimensional surface at $r = r_3 + \delta$, with
the limit $\delta \to 0$ taken at the end. Notice that total
derivatives in the bulk, such as $\Box \phi$, do {\it not} give
boundary contributions at the position of the 4-brane because the
4-brane does not represent a boundary of the manifold $M$.
(Alternatively, the boundary contributions cancel for each of the
two copies of $\widehat M$ of which $M$ is composed.)

We see that for the $N = 4^g$ theory, where $\zeta = -1$, the bulk
contribution to the result cancels, leaving only terms evaluated
at the positions of the two branes.

It is straightforward to see that the 3-brane contributions to
eq.~\pref{RemoveB} vanish for the warped solution considered
earlier. The first term does so straightforwardly because of the
dilaton 3-brane boundary condition, which required us to choose
$\lambda_3 = 0$. The surface term also vanishes in this case
because $B^{MN} = 0$ and because the outward-pointing unit normal,
$n_M dy^M = - dr/\sqrt{g^{rr}}$, contributes an amount $n_M
\partial^M \phi = - \sqrt{g^{rr}} \, \phi' = \sqrt{ab} \, \phi'$,
which vanishes as $r \to r_3$ by virtue of the vanishing of
$b(r_3)$. This agrees with the general scaling argument given
above, which indicated self-tuning in the case $\lambda_3 = 0$.

We see that the 4-brane contribution to $\rho_{\rm eff}$ also
vanishes provided that $\lambda_4 = \frac12$, again in
agreement with the general scaling argument. Since our explicit
warped solution of earlier sections does not require a specific
value for $\lambda_4$, we are free to make this choice and so to
ensure the vanishing of $\rho_{\rm eff}$. It is a straightforward
exercise to verify that the choice $\zeta_4 = - \, \frac12$ would
also be required to ensure $\rho_{\rm eff} = 0$ if we had taken $q
\ne 0$ and followed the 4-brane fields $\sigma$.

\subsection{Salam-Sezgin Supergravity}
We here repeat the above exercise for the solution to Salam-Sezgin
supergravity. We keep the presentation concise since the
arguments largely follow the discussion in
ref.~\cite{branesphere}.

For two parallel 3-branes positioned at $y = y^m_\pm$ in the
internal dimensions the effective 4D vacuum energy in Salam-Sezgin
supergravity is
\eqa \label{rhoeffSS}
    \rho_{\rm eff} &=& \sum_{i=\pm} w^2(r_i)\, T_i + \int_M d^2y \; e_2 \,
    w^2
    \left[\hf R_6 + \hf
    (\partial \phi)^2 + \hf G_{ab} (D \Phi^a) (D \Phi^b)  \right. \nn\\
    && \qquad \left. \left. + \frac{1}{12} \, e^{-2\phi} \, G^2
    + \frac14 \, e^{-\phi} \, F^2 + v(\Phi) \, e^\phi
    \right]_{cl} \right|_{{g_{\mu\nu} =
    \eta_{\mu\nu}}}
\eeqa
where $w(r) = 2\, r$ is the warp factor, and $M$ denotes the
internal two-dimensional bulk manifold. As before the subscript
`$cl$' indicates the evaluation of the result at the solution to
the classical equations of motion.

Using the Einstein equation to eliminate the metric gives
\eq
    R_6 = - (\partial \phi)^2 - G_{ab} D\Phi^a D\Phi^b -
    3 v(\Phi) \, e^\phi - \frac14 \,
    e^{-\phi} \, F^2 - {2 \over e_2} \, \, \sum_{i=\pm} T_i \,
    \delta^2(y-y_i)\,  ,
\eeq
and using this in $\rho_{\rm eff}$ gives
\eq \label{RemoveGSS}
    \rho_{\rm eff} = \left. \int_M d^2y \; e_2 \, w^2 \left[  \frac{1}{12}
     \, e^{-2\phi} \, G^2 + \frac18 \,
    e^{-\phi} \, F^2 - \frac12 \, v(\Phi) \, e^\phi
    \right]_{cl} \right|_{cl} \, .
\eeq

The dilaton equation of motion now reads
\eq
    v(\Phi) \, e^\phi - \sfrac14 \, e^{-\phi} \, F^2 - \,
    \sfrac16 \, e^{-2\phi} \, G^2 = \Box \phi - \sum_{i = \pm}
    \lambda_i T_i e^{\lambda_i \phi} \, \frac{1}{e_2} \,
    \delta^2(y - y_\pm) \, ,
\eeq
which gives when inserted into eq.~\pref{RemoveGSS}
\eqa \label{rhoeffresult}
    \rho_{\rm eff} &=&  -\, \frac12 \int_M d^2y \; e_2 \,w^2 \,
     \Box \phi_{cl}  + \frac{1}{2}\, \sum_{i=\pm} \lambda_i T_i \,
     w^2(r_i) \nn\\
    &=& \sum_{i=\pm} w^2(r_i) \,\Bigl[ \frac{1}{2} \lambda_i T_i
    - \pi \, e_2
    \, n_M \, \partial^M \phi
     \Bigr]_{r = r_i} \, ,
\eeqa
where we evaluate total derivative using the boundary surface
$\partial M_i$, consisting of an infinitesimal region surrounding
the 3-brane positions. For the solution considered above this
consists of an infinitesimal circle surrounding the brane
positions at $r = r_\pm$.

The two contributions to $\rho_{\rm eff}$ therefore vanish when
evaluated at the solutions derived in earlier sections. The first
term vanishes because we have already seen that the solution
described above requires $\lambda_\pm = 0$, and the second 
likewise vanishes
because $\phi'$ is bounded but $n_M \, \partial^M \phi =
\sqrt{g^{rr}} \, \phi' = \sqrt{h}\, \phi'$ vanishes at the brane
positions, $r = r_\pm$.

\section{Discussion}
In this paper we constructed explicit warped, axisymmetric solutions
to the dilaton-Einstein-Maxwell field equations arising from both
Romans' and Salam-Sezgin supergravity in six dimensions. We
identified the circumstances under which they may be interpreted as
being generated by simple 3- and 4-brane sources, and what
geometrical features are required in order for the resulting brane
systems to be used as brane-world models having a realistic
electroweak hierarchy. Since all of the solutions have flat
4-dimensional sections regardless of the values of the tensions
and couplings on the various branes, they resemble the unwarped
solution of ref.~\cite{branesphere}. We therefore also examine
more generally how self-tuning of the 4D cosmological constant
arises in these models. This allows us to identify some of the
issues which must be addressed in order to promote these features
into a real solution to the cosmological constant problem.

Our results, in more detail, are as follows.

\subsection{Brane World Solutions}
We considered two kinds of supergravities --- Romans' and
Salam-Sezgin --- whose bosonic parts mainly differ in the overall
sign of the exponential potential they predict for the 6D dilaton,
and we found warped solutions for both theories.

\vfill\eject

\ssubsubsection{Romans' Supergravity}
The warped solution to Romans' supergravity can have a conical
defect at its origin, which we interpreted as the position of a
3-brane. It is also terminated by a boundary 4-brane in order to
ensure the transverse dimensions to have finite volume. By making
simple assumptions about the physics of the 3- and 4-brane, we
investigated how the parameters of the bulk solution are related to
the physical properties of the branes. We found that in the generic
case the number of boundary conditions is larger than the number
of integration constants, implying that the bulk solutions we find
can only be interpreted as being generated by the assumed brane
sources if some of the brane couplings are adjusted.

In detail, our assumption that the dilaton remains nonsingular at
the 3-brane position required us to choose the dilaton 3-brane
coupling to vanish: $\lambda_3 = 0$. In addition, we found that the
magnetic flux is determined both by the metric/dilaton boundary
conditions and by a topological condition. These conditions are
generically not consistent with one another, but can be made consistent by
adjusting the background gauge coupling, $g$. Alternatively, since
we assumed for simplicity the 4-brane to be superconducting, we could
satisfy this equation by adjusting the 4-brane symmetry breaking
scale (or `penetration depth'), $q$.

In the generic case, the 4-brane couplings break the classical
scale invariance of the bulk theory and so we were able to determine all
parameters of the solution using the boundary conditions. In this
sense our ansatz has no moduli, and so does not have a
classically-massless dilaton or breathing mode. The scale
invariance is not broken for the special case $\lambda_4 =
\frac12$ and $\zeta_4 = - \frac12$, and in this case there is at
least one flat direction.

We briefly examined brane-world models based on this solution and
saw that an electroweak hierarchy could be obtained, but only by
choosing a hierarchy in the underlying 6D theory or by adjusting
dilaton couplings to be near to their scale-invariant values. In
the precisely scale-invariant case the overall hierarchy could be
simply set by the position chosen along the flat direction, and so
would have to be explained by whatever physics stabilizes this
direction.

Our inability to account for the electroweak hierarchy as cleanly
as was possible for nonsupersymmetric systems follows from the
presence of the dilaton, since the dilaton is free to roll to its
potential minimum asymptotically, at which point the spacetime
curvature also vanishes. Consequently the solution we found is
asymptotically locally flat (conical), rather than being
asymptotically anti-de Sitter, as is the case for the 5D Randall
Sundrum \cite{RS} and 6D ADS soliton \cite{AdSsoliton,6DRS,OurADS}
solutions; in the latter the dilaton is replaced by a negative
cosmological constant.

The warping of the metric ensures that the KK spectrum of the model
need not involve many states lighter than the weak scale even if
the proper radius of the internal space is comparatively large.
This is because the lightest bulk KK modes tend to be localized
near the 3-brane, and so do not `see' the entire extent of the
extra dimensions. In particular, many of the attractive
relationships between scales which occur in the unwarped case
(such as that relating the electroweak hierarchy to the effective
cosmological constant) do not appear to also hold for these warped
solutions.

\ssubsubsection{Salam-Sezgin Supergravity}
The warped solution to Salam-Sezgin supergravity can have either
one or two conical defects, which we interpreted as the position of
one or two 3-branes. Again the number of boundary conditions is
larger than the number of integration constants, and so the bulk
solutions are only produced by the assumed branes if their
couplings are adjusted in particular ways.

As for the Romans' case, nonsingularity of the dilaton requires
vanishing dilaton 3-brane couplings: $\lambda_\pm = 0$.
Furthermore, the 3-brane tensions are subject to a topological
condition which generalizes the condition found in the unwarped
case (for which the tensions must be equal). Finally, we found a
topological condition on the total magnetic flux through the
space, whose satisfaction requires the adjustment of one of the
couplings, such as the background gauge coupling, $g$.

Since the required dilaton couplings preserve the classical scale
invariance of the bulk theory there is at least one classically
flat direction corresponding to the overall volume of the internal
dimensions.

Brane-world models based on this solution can have acceptable
electroweak hierarchies, but apparently only by inserting the
required hierarchies by hand into the 6D theory. Again, this can
be chosen to be along the flat direction, pending an understanding
of modulus stabilization in this direction. Unlike the unwarped
example, there does not seem to be any compelling numerology which
relates the required extra-dimensional sizes to the observed
electroweak or cosmological constant hierarchies.

\subsection{Self-Tuning Issues}
All of the solutions which we considered have flat 4D slices
for any values
of the various couplings, suggesting these share the self-tuning
properties of the unwarped example. In section 4 we traced the
origin of self-tuning to the classical bulk scale invariance, and
so made an explicit connection between our higher-dimensional
self-tuning and Weinberg's general formulation of self-tuning in
four dimensions. This connection allowed us to clarify how the
usual objections to self-tuning arise in the 6D context.

\ssubsubsection{Tuning of Couplings}
Since special adjustments of couplings are required to interpret
our solutions as the fields set up by simple brane sources, one
worries that these adjustments may also be responsible for tuning
the 4D cosmological constant. But given that classical scale
invariance is the central property required, we believe the
dilaton/brane coupling conditions ($\lambda_3 = 0$ for 3-branes
and $\lambda_4 = - \zeta_4 = \frac12$ for 4-branes) are the ones
which are important for the self-tuning mechanism (as was also
true in the original self-tuning solutions of \cite{adks}).
 Phrased in this
way, the self-tuning is not seen as an exclusively 6D property,
and is likely to apply more generally to brane configurations in
compactified spaces.

On the other hand, since our self-tuning calculations of section 4
do not use any of the detailed properties of the solution, we
believe the topological conditions which our models satisfy do not
play a similarly important role. For instance, for 3-branes we
find that the necessary condition for a solution to be self-tuning is 
that it have a nonsingular dilaton at the brane positions. 
Given only this,
the bulk curvature automatically cancels the brane tensions
regardless of the values these tensions take. In particular, the
cancellation occurs for {\it any} value of the tensions, and does not depend on
whether the tensions are related to one another by topological
conditions.

We believe the same to be true for the magnetic flux-quantization
conditions, since these conditions are actually very similar in
form to the tension constraints. To see this, imagine including
the following direct 3-brane coupling to the magnetic flux,
obtained by integrating the Hodge dual ${}^*F$ over the
four-dimensional brane world volume
\eq \label{HodgeAction}
    \Delta S_3 = -\, \frac{\Q}{2} \int_{r_\pm} d^4\xi \,
    e_4 \, e^{-\phi} \, \epsilon^{m n}
    F_{mn}
    \, .
\eeq

This term causes the 3-brane itself to carry magnetic flux, since
it causes the flux through an infinitesimal surface surrounding
the brane to be nonvanishing. This may be seen from the Maxwell
field equation, which is modified to become (when $B_{MN} = 0$)
\eq
    \partial_M \Bigl( e_6 \, e^{-\phi} \, F^{MN} \Bigr) =
    \delta_n^N
     \Q \, \partial_m \Bigl( e_4 \, e^{-\phi} \, \epsilon^{mn}
    \delta^2(\vec{y} - \vec{y}_3) \Bigr) \, ,
\eeq
thus showing that the bulk magnetic flux acquires delta-function
contributions at either of the two brane positions.

Given this choice, for the Salam-Sezgin model the gauge potential
again has the form of eq.~\pref{Apatch}, but now with the
condition that $A \to (\Q_\pm/2\pi) \, \exd \theta$ as $r \to
r_\pm$, leading to
\eq \label{Apatcha}
    A_\pm = \left[ \frac{\Q_\pm}{2\pi} + \frac{{\cal A}}{2 \, r_\pm^2}
    \left( 1 - \frac{r_\pm^2}{r^2} \right) \right] \exd \theta \,,
\eeq
in the patch centered at $r = r_\pm$. Requiring, as before, the
two patches to be related by a periodic gauge transformation in
this case replaces eq.~\pref{topcond} with the condition
\eq
    \frac{\Q_+ - \Q_-}{2\pi} + \frac{{\cal A}}{2} \left(
    \frac{1}{r_+^2} - \frac{1}{r_-^2} \right) = \frac{n}{g}
    \, .
\eeq
This is the direct analogue of the tension constraint,
eq.~\pref{Tconstraint}, which relates the 3-brane tensions for
compact extra dimensions. (No integer appears in the tension
constraint because the integer has already been chosen, since the
internal space has Euler number 2.)

Seen in this light, the flux-quantization constraint may be viewed
as a condition on the charges carried by various branes, rather
than as a tuning of gauge couplings which are parameters of the
bulk action. (Constraints on the gauge couplings, like
eq.~\pref{topcond}, are seen in this light as being forced by the
specialization to two branes having identical flux: $\Q_+ =
\Q_-$.) In this sense, the tuning simply expresses global
constraints on what combinations of brane charges it makes sense
to include within extra dimensions of the assumed topology, in a
similar manner to the well-known Gauss' Law requirement that the
total charge for a collection of particles in a compact space must
vanish.

Just like the Gauss' Law constraint (or the quantization condition
for a magnetic monopole, or the flux-quantization condition for
annular superconductors) we expect these constraints to be
stable under UV-sensitive radiative corrections in the 6D theory;
hence they are not fine-tunings in the sense of the
cosmological constant problem.
This radiative stability relies on the fact that short-distance
quantum corrections must be local, and so are unlikely to affect
long-distance topological effects.

Of course, an explicit demonstration of this stability is more
persuasive than a hand-waving argument in its favor, and work on 
this is in progress.

\ssubsubsection{Quantum Corrections and the No-Go Theorem}
Because models of this class obtain a zero 4D cosmological
constant by virtue of their classical scale invariance, they fall
directly into the category of self-tuning models, and so also
into Weinberg's related no-go theorem, described in
ref.~\cite{CCReview}. This suggests that there are {\it two} ways
in which quantum corrections can ruin the self-tuning,
rather than one.

The simplest problem which quantum effects raise is that they need
not respect the scale invariance. This is certainly true in the
theories studied here, for which the scale transformations do not
leave the action invariant, but rather transform it into a
multiple of itself. Although this suffices to ensure a symmetry of
the classical equations of motion, it does not guarantee
invariance for the full path integral and so the low-energy
quantum-corrected action need not be scale invariant.

Even if quantum corrections were to respect scale invariance, the
no-go theorem of ref.~\cite{CCReview} raises another problem. This
is because within any phenomenologically-successful
scale-invariant theory the symmetry must be spontaneously broken
in order to allow nonzero particle masses. It must therefore
contain an effective 4D dilaton, $\varphi$, which is the 4D
Goldstone boson for the scale invariance, and which therefore shifts
under a scaling transformation: $\varphi \to \varphi + c$. (All
other fields can then be made invariant by performing appropriate
field redefinitions \cite{PhysRep}.) This transformation law
ensures that the dilaton equations of motion suffice to ensure that
flat space solves Einstein's equations. In the 6D models studied
here, this dilaton is a linear combination of the 6D dilaton,
$\phi$, and the internal metric's `breathing' mode.

The difficulty with these models is that scale invariance cannot
forbid a term in the 4D scalar potential of the form $V_{\rm eff}
= v \, e^{a \varphi}$, where $v$ and $a$ are dilaton-independent
quantities. Since these are not constrained at all by scale
invariance, scale invariance by itself cannot ensure $v = 0$, and so
typically quantum corrections make $v \ne 0$ even if they are
scale invariant. This lifts the degeneracy along the $\varphi$
direction, making the ground state unique, and thereby leads to
a vacuum which does not spontaneously break scale invariance at
all. Although the 4D cosmological constant vanishes, it does so by
driving the theory to a scale-invariant vacuum. The resulting
theory cannot be said to solve the cosmological constant problem,
because it is not a great achievement to obtain a vanishing
cosmological constant in a theory for which all masses are also
zero.

There are clearly two problems, and supersymmetry may be able to
help with both of them. Although quantum corrections do break
scale invariance, and can lift flat directions, supersymmetry
typically ensures the scale for doing so is the
supersymmetry-breaking scale. In the 6D models of interest here
the self tuning mechanism can handle any quantum corrections on
the brane, but cannot do so at the quantum level for the bulk
modes. However the scale of supersymmetry breaking for the dilaton
sector in these models is of order the bulk KK mass scale,
$m_{KK}$, which can be much smaller than the usually-assumed TeV
scale without running into observational difficulties. This is
particularly striking for the unwarped solutions, for which
$m_{KK}$ can be as small as $10^{-3}$ eV.

Although no general proof exists that quantum corrections to the
dilaton potential must be as small as $O(m_{KK}^4)$, there are
encouraging indications. Explicit calculations in (unwarped)
supersymmetric string and field theories with supersymmetry broken
on branes indicate that the effective 4D cosmological constant
generated at one loop {\it are} of order $m_{KK}^4$, rather than
being set by the scale of the brane tension \cite{CasimirCalcs}.
In six dimensions self-tuning itself has been argued for unwarped
geometries to ensure that quantum corrections to the dilaton
potential are at most of order $M_w^2 m_{KK}^2$
\cite{branesphere}.

Notice that $m_{KK}$ is typically much smaller in the unwarped
compactifications than is found for the solutions examined here.
Consequently it is the unwarped, large-extra-dimensional scenario
which is the most attractive for potentially addressing the
cosmological-constant problem in the low-energy 4D theory.
Furthermore, the choice of the unwarped vacuum is likely to be
stable against quantum corrections because (unlike the warped
solutions) in the absence of branes it preserves an unbroken $N=1$
supersymmetry in four dimensions.

Although none of these lines of argument are yet conclusive, we
believe it is sufficiently encouraging to warrant more fully
exploring how quantum corrections arise in the low-energy sector
of these theories.

\vfill\eject

\subsection{Open Issues}
Our discussion suggests several directions for further
exploration. Most notable among these is the solution to the
general problem of finding the back reaction of simple 3-brane
configurations in six dimensions without the neglect of dilaton or
electromagnetic couplings. Given the general configuration it
would be possible to identify whether the brane-coupling choices
we make play an important role in the low-energy properties and
with the self-tuning of the 4D cosmological constant.

An equally important issue to be addressed is the extent to which
bulk radiative corrections change our results. In particular one
would like to address the extent to which supersymmetry helps
protect the electroweak hierarchy and 4D cosmological constant,
given that these are chosen to be acceptably small at the
classical level.

Given that the field equations we examine are supersymmetric, it
would be useful to know how our solutions may be embedded into a
still-higher-dimensional theory like 10D supergravity or string
theory. At present this connection can be made more explicit for
Romans' supergravity --- such as for the explicit lift to ten
dimensions described in the appendix (section 6) --- because it is known how to obtain
this theory by consistent truncation from higher dimensions.
Similar constructions for Salam-Sezgin supergravity are presently
being developed, \cite{newsugra},
\cite{paulired}.\footnote{
Ref. \cite{paulired} obtains an embedding of Salam-Sezgin supergravity
by performing a consistent 
Pauli reduction of 11D/10D supergravity on the non-compact
hyperboloid ${\cal H}^{2,2}$ times $S^1$.} 

We believe that further explorations in these directions is
warranted by the preliminary features we have been able to identify
here.

\acknowledgments We thank G. Gibbons and C. Pope for stimulating
conversations on the Salam-Sezgin model.
 Y.A., H.F. and C.B.'s research is partially funded
by grants from McGill University, N.S.E.R.C. (Canada) and F.C.A.R.
(Qu\'ebec). S.P. and  F.Q. are partially supported by PPARC.
 G.~T.~is supported by the European TMR Networks HPRN-CT-2000-00131,
HPRN-CT-2000-00148 and HPRN-CT-2000-00152. I.~Z.~ is
supported by CONACyT, Mexico. F.Q.
thanks the Physics Department of McGill University for hospitality
during the late stages of this project. C.B. and F.Q. thank Second
Cup for extensive use of their facilities.

\section{Appendix: Romans with 10D Lifts}
Much of the motivation for studying compactifications of
higher-dimensional supergravities comes from their interpretation
as low-energy vacua of string theory. This allows the
identification of any phenomenologically-attractive low-energy
features of these models to be taken as guidelines when searching
for realistic string vacua. For these purposes it is necessary to
know how a lower-dimensional supergravity arises from its
higher-dimensional counterparts in order to be able to properly
identify how the lower-dimensional fields correspond to explicit
string modes. Since this matching has been partially performed for
the supergravities we use, we pause here to record how it works.
Our discussion, given for the $N = \tilde{4}^g$ theory (but
similar to what happens for $N = 4^g$), follows that given in
ref.~\cite{NPST}.

\smallskip
Given a solution to Romans' $N = \tilde 4^g$ six-dimensional
supergravity involving the fields we consider here, one may always
generate a solution to the bosonic field equations of a
10-dimensional supergravity, in which the metric, dilaton and
Ramond-Ramond (RR) 4-form, $F_{\it 4}$, take nontrivial values. We
will consider, as internal manifold in the uplifting procedure,
the space $S^{3} \times T^{1}$.

If we adopt a notation for which 10D and 6D quantities are
distinguished by marking the 10D fields using tildes, then the
same uplifting procedure gives us the relevant part of the bosonic
action for the 10D theory, that may be written as
\eq\label{tendact2a} {\mathcal L_{10}} = -\, \tilde R -\,
\frac{1}{2} (\partial\tilde{\phi})^2 -\frac{1}{2}
e^{-\tilde\phi/2} \tilde F_{\mu\nu\rho\lambda}\tilde
F^{\mu\nu\rho\lambda} \,. \eeq
We adopt the standard convention that $e^\phi \to 0$ corresponds
to weak string coupling, so our results differ from the
conventional form for the (truncated) bosonic action of 10D type
IIA supergravity by simply re-defining the scalar field according
to $\tilde{\phi} \to -\tilde{\phi}$ \ \cite{NPST}.

The ten dimensional field configuration corresponding to a
solution for the equations relative to (\ref{tendact2a}) is then
given in terms of the six-dimensional one by the following
expressions:
\eqa\label{tensix}
    d\tilde s^2_{10}& =& \frac{1}{2}e^{{\phi}/{4}}
    ds^2_6 + \frac{1}{2\,g_2^2} e^{{-3\phi}/{4}} \sum_{\alpha=1}^3 \left(
    \sigma^\alpha - \frac{g_2\, A^\alpha_1}{\sqrt{2} } \right)^2
     + e^{{5\phi}/{4}} dZ^2\,, \nonumber\\
    \tilde F_{\it 4} &=& \left( \tilde{G}_{\it 3} -
    \frac{1}{\sqrt{2}\,g_2^2} \, h^1\wedge h^2 \wedge h^3
      + \frac{1}{2\, g_2}F^\alpha_2 \wedge h^\alpha\right) \wedge \exd Z \,,
    \label{fourform} \\
    \tilde\phi &= &\frac{1}{2} \, \phi\,, \nonumber
\eeqa
where $\theta_1, \psi, \varphi$ and $Z$ are coordinates on the 4
new dimensions, $h^\alpha = \sigma^\alpha - \frac{ g_2}{\sqrt{2}} \, 
A^\alpha_1$,
and the $\sigma^\alpha$ are left-invariant 1-forms for $SU(2)$ given by
\eqa\label{sigmas}
    && \sigma_1 = \cos\psi \, \exd\theta_1 + \sin\psi \sin\theta_{1} \,
    \exd\varphi \,, \nn \\ 
    &&\sigma_2 = \sin\psi \, \exd\theta_1 - \cos\psi \sin\theta_{1}
    \, \exd\varphi \,, \\ 
    && \sigma_3 = \exd\psi + \cos\theta_1 \, \exd\varphi \nn \,.
\eeqa
The 3-form, $\tilde{G}_{\it 3}$, appearing within the expression
for the 4-form in eq.~(\ref{fourform}) is the 6-dimensional dual
\eq\label{f3}
    \tilde{G}_{\mu\nu\rho} = \frac{1}{6} \, e^{2 \phi}\,
    \epsilon_{\mu\nu\rho\lambda\beta\gamma}G^{\lambda\beta\gamma}\,,
\eeq
of the 3-form field strength of the field $B$ appearing in the 6D
Romans' Lagrangian.

The obtained uplifted solution is quite interesting, since, in
general, it can geometrically be interpreted as a configuration of
intersecting D-branes. We see in the following how this works in a
specific example.

We now specialize these general formulae to specific 6D solution
considered in previous sections into a ten dimensional solution to
the equations obtained from the Lagrangian (\ref{tendact2a}). We
get
\eqa
    d\tilde s^2_{10}& =& \frac{e^{\phi_0} \, r^{1/4}}{2 }
    \left(a_0\, r \,\eta_{\mu\nu} \exd x^\mu \exd x^\nu + a_0\,r\,
    b(r) \, \exd\theta^2
    +\frac{\exd r^2}{a_0\,r\, b(r)} \right) \nonumber \\
     && \hskip1cm +\frac{e^{-3\phi_0/4}}{2\,g_2^2\,r^{3/4}}
    \left((\sigma^1)^2
    +(\sigma^2)^2 + \left(\sigma^3
    +\frac{g_2\,A\, e^{\phi_{0}}}{2 \sqrt{2}\,a_{0}^{2}\,
     r^{2}} \right)^2 \right)
    + e^{5 \phi_0/4} r^{5/4} \exd Z^2\,,\nn \\
    \tilde\phi &= &\frac{\phi_0}{2} + \frac{1}{2}\ln {r}\,, \\
    \tilde F_{\it 4} &=& \left(- \frac{1}{g_2^2\sqrt{2}} \sin\theta_1
    \, \exd\theta_1 \wedge \exd\varphi \wedge \exd\psi
    - \frac{A e^{\phi_{0}}}{4\,g_2\,a_{0}^{2}\,r^{2}}
     \sin\theta_1\, \exd\varphi  \wedge \exd
                               \theta_1\wedge \exd \theta \right.\nonumber \\
&& \qquad \qquad \qquad \left.
    + \frac{A\,e^{\phi_0}}{2\,a_0^2\, g_2\, r^3}\, \exd r\wedge \exd
    \theta \wedge (\exd\psi + \cos\theta_1 \, \exd\varphi) \right)
    \wedge \exd Z \,. \nn
\eeqa

This configuration represents three D4 branes that intersect over
three spatial directions. Two of the D4 branes, moreover,  wrap
the three sphere, which is part of the internal manifold. As one
might expect, in the supersymmetric limit discussed in the
previous section, the angle of intersection vanishes.

\section{Appendix: Explicit Solution with $\lambda_4 = - \zeta_4 = \frac12$}

Here we will show explicitly how the counting of free parameters
works in the Romans case. In general we have 5 parameters $A, B,
C, \alpha_3, r_4$. The parameter $C$ can be eliminated right away
using the matching of the gauge potentials as in equation
(\ref{max4b}). So we will concentrate on the $4$ parameters $A, B,
\alpha_3$ and $x={r_3}/{r_4}$. To determine them we have one
condition coming from the 3-brane, namely, eq.~ (\ref{met3}), plus
4 conditions coming from the 4-brane, which are eqs.~(\ref{EDeqn},
\ref{Deqn}, \ref{ED2eqn}) and the flux quantisation condition
(\ref{fluxquantization}). So in the general case we have one more
equation than parameters and therefore there has to be at least
one constraint
 involving $g_2$ and the  brane parameters $T_4, T_3, q, \lambda_4, \zeta_4$.

In the conformal invariant case $\lambda_4=-\zeta_4=1/2$, equation
\ (\ref{ED2eqn})\ is automatically satisfied. We could have then
concluded that
 with one less equation we have the same number of equations and paramters
 and no constraint may be
needed. However this is not the case. The reason is precisely
because  in this case we have the extra scaling symmetry
(\ref{rescaling2}) which implies that one of the parameters is
actually redundant. We can see this explicitly by trying to solve
the 5 equations mentioned above. This is what we will do now.

First we need to recall  that
 $e^{\phi(r_4)}=\alpha_3 x$ and
$a(r_4)=1/x$ as well as the relation between $A_\theta$ and $A$
given in (\ref{max4b}). Also since $r_3$ appears explicitly in
most of the equations we need to use often the expression
(\ref{rthree2}). To simplify the calculations we work in the limit
of large $r_4$ meaning that $b(r_4)\sim b_1=g_2^2\alpha_3 r_3^2/4$
and $b'(r_4)\sim 2B/r_4^3$.

From this we can see that equations (\ref{EDeqn}) and (\ref{Deqn})
can be solved for $x$ and $A$ giving:
\eq x^2\ = \ \rho\left[-1\pm \sqrt{1-\frac{\mu}{\rho}}\ \right]
\eeq
where $\rho\equiv 2(T_4-g_2)$ and $\mu= q^2/g_2^2$. For $A$ we
find:
\eq A^2(x)\ = \ \left(1+\frac{2g_2^2}{q^2} x^2\right)^{-1} \eeq
For the remaining parameters $\alpha_3$ and $B$, we can easily see
that equation~(\ref{fluxquantization}) implies:
\eq \alpha_3 r_3\ = \ \frac{2N}{gg_2
A}\left[1-x^2\left(1-\frac{g_2}{q^2}\right)\right]^{-1}\ \equiv
F_1(x)\, \eeq
and the 3-brane condition (\ref{met3}), implies
\eq \frac{B}{r_3^3}\ = \
\frac{\left(1-\frac{T_3}{2\pi}\right)\left(1-A^2\right)}{1+A^2}\
\equiv F_2(x) . \eeq
Since $r_3$ appears on both equations we can eliminate it by
taking their ratio. But precisely the ratio in the left hand side
 is what appears in the expression for $r_3$ in (\ref{rthree2}).
This then implies that:
\eq \frac{F_1(x)}{F_2(x)}\ = \ \frac{4}{g_2^2} \frac{1}{1-A^2(x)}
\eeq
This is  a constraint that involves $only$ the external
parameters: $T_3, T_4, q$ as well as $g_2$ (since we have the
explicit solutions for $x$ and $A$) but not $\alpha_3, B$. This
also implies that one combination of the parameters $\alpha_3, B$
remains unfixed. Therefore we have shown explicitly that in this
case there is still one free parameter, unlike the generic
nonconformal, cases and that, similar to those cases, there is
still one consistency constraint to be satisfied. This illustrates
the general arguments given in the text.

Finally we can see from the expression for $x$ above, which
amounts to fixing the size of the extra dimensions,
 that in order to obtain a hierarchy $x\ll 1$ we may
have to have either $\rho\ll 1$ or $q^2\ll g_2^2$.


\end{document}